\documentclass[article,12pt]{elsarticle}
\usepackage[a4paper,text={400pt,650pt},centering]{geometry}
\usepackage[utf8]{inputenc}
\usepackage{amsmath}
\usepackage{relsize}
\usepackage{verbatim}
\usepackage{appendix}
\usepackage{mathtools}
\usepackage{tikz}
\usetikzlibrary{decorations.markings}
\usepackage{braket}
\usepackage[new]{old-arrows}

\usepackage{amsmath,bm}
\usepackage{enumerate}
\usepackage{pgfplots}
\usepackage{fancyhdr}
\usepgfplotslibrary{fillbetween}
\usetikzlibrary{patterns}
\pgfplotsset{compat=1.12}
\usepackage{mathtools}
\usepackage{graphicx}
\usepackage{caption}
\usepackage{subcaption}

\usepackage{graphicx}
\usepackage[compat=1.0.0]{tikz-feynman}
\usepackage{amsfonts,amssymb,amsthm,epsfig,epstopdf,url,array}
\def\ov{\overline}
\usepackage{braket}

\usepackage{enumerate}

\usepackage{lineno,hyperref}
\modulolinenumbers[5]

\journal{A}
\biboptions{sort&compress}   %% from Manual

\let\today\relax
\makeatletter
\def\ps@pprintTitle{%
    \let\@oddhead\@empty
    \let\@evenhead\@empty
    \def\@oddfoot{\footnotesize\itshape
         {\sc MPP--2022--001} \hfill\today}%
    \let\@evenfoot\@oddfoot
    }
\makeatother

\def\Tr{\mathop{\rm Tr}}

\theoremstyle{remark}

\newcommand{\pf}{\mathrm{Pf}}

\newcommand{\sect}[1]{ \section{#1} \setcounter{equation}{0} }
\newcommand{\subsect}{\subsection}
\newcommand{\req}[1]{(\ref{#1})}
\def\vev#1{\langle #1 \rangle}

\def\fc#1#2{\frac{#1}{#2}}
\def\h{\frac{1}{2}}
\newcommand{\nwc}{\newcommand}
\nwc{\ba}  {\begin{array}}
\nwc{\ea}  {\end{array}}
\nwc{\bdm} {\begin{displaymath}}
\nwc{\edm} {\end{displaymath}}

\nwc{\bea} {\begin{equation}\ba{lcl}}
\nwc{\eea} {\ea\end{equation}}

\nwc{\be} {\begin{equation}}
\nwc{\ee} {\end{equation}}

\nwc{\bda} {\bdm\ba{lcl}}
\nwc{\eda} {\ea\edm}

\nwc{\bc}  {\begin{center}}
\nwc{\ec}  {\end{center}}

\nwc{\ds}  {\displaystyle}

\nwc{\nn} {\nonumber}
\nwc{\nnn} {\nonumber \vspace{.2cm} \\ }
\nwc{\ra}{\rightarrow}
\nwc{\lra}{\longrightarrow}
\def\lf{\left}\def\ri{\right}

\nwc{\p} {\partial}
\def\IR{{\bf R}}

\def\ap{\alpha'}

\def\ov{\overline}
\def\th{\theta}

\def\eps{\epsilon}
\def\z{\zeta}

\def\Oc{{\cal O}}

\def\IR{{\bf R}}
\def\IC{{\bf C}}
\def\IP{{\bf P}}

\def\Ac{{\cal A}}

\def\Sc{{\cal S}}

\def\th{\theta}
\def\eps{\epsilon}

\def\si{\sigma}\def\Si{{\Sigma}}
\def\om{\omega}

\begin{document}

\begin{frontmatter}

\title{{\sc Einstein  Yang-Mills Amplitudes  \\ from Intersections of Twisted Forms}\vskip1.5cm}

\author{Pouria Mazloumi\ and\  Stephan Stieberger}


\address{Max-Planck Institut f\"ur Physik\\Werner–Heisenberg–Institut, 80805 Munich, Germany}

\address{{\it Emails:} pmazlomi@mpp.mpg.de, stieberg@mpp.mpg.de \vskip2cm}

\begin{abstract}
We present a geometric derivation of all--multiplicity (single--trace) tree--level Einstein Yang-Mills (EYM)  amplitudes ${\cal A}(n;r)$ involving  $n$ gluons and $r$ gravitons by a bilinear of two twisted differential forms on the moduli space of Riemann spheres with $m\!:=\!n\!+\!r$ punctures. The differential forms are gained by studying the underlying superstring disk amplitude
and proposing an embedding of the disk onto the sphere. This map can be interpreted as a geometrical map from the open superstring  to a heterotic or ambitwistor string structure.
Then, the twisted intersection number  of the two $m$--forms, which is obtained by integrating over the moduli space of Riemann sphere with $m$ punctures, reproduces in the infinite inverse string tension limit $\alpha'\!\rightarrow\! \infty$ the corresponding CHY formula of the EYM amplitude. 
To bolster our findings we study the disk amplitude of open and closed strings  using the Gra\ss mann description of the underlying superstring amplitude, map it to a closed string amplitude and consider the $\alpha'\!\rightarrow\! \infty$ limit. Finally, we present an all--multiplicity decomposition formula of any EYM amplitude ${\cal A}(n;r)$ as linear combination over $(m\!-\!3)!$ pure $m$ gluon subamplitudes.
\ \\

 \end{abstract}


\end{frontmatter}

\newpage

\setcounter{tocdepth}{2}
\tableofcontents
\newpage
\section{Introduction}

Amplitudes are traditionally formulated in four--dimensional momentum space  where translation symmetry is manifest. However, not all symmetries are manifest in this formulation. 
The study of hidden symmetries  of scattering amplitudes  has revealed deep connections between gauge and gravitational interactions established  in striking identities such as 
Kawai--Lewellen--Tye (KLT) \cite{klt} and Bern--Carrasco--Johansson (BCJ) \cite{BCJ1} equations 
 - manifested also in double--copy--constructions \cite{Bern:2019prr}.
Currently, there is an intense global research activity to understand the origin of these relations.   Furthermore, based on the works \cite{Stieberger:2014cea,Stieberger:2015qja,Stieberger:2015kia,Stieberger:2015vya,ST} on Einstein--Yang--Mills (EYM) amplitudes and their relations to pure gluon amplitudes there has been growing interest in the study of their underlying structure \cite{p,delaCruz:2016gnm,Schlotterer:2016cxa,Fu:2017uzt,Chiodaroli:2017ngp,Teng:2017tbo,Roehrig:2017wvh,Du:2017gnh}, cf. also some earlier work  \cite{Bern:1999bx,Selivanov:1997aq,Selivanov:1997ts}.
These connections  within and between gravity and gauge theory scattering amplitudes ask for a unification between these theories of the sort inherent to some more fundamental formulation or dual description of interactions.

A natural geometric framework for amplitudes is provided by  the geometry of Riemann surfaces (a two--dimensional complex manifold) with marked points.
Riemann surfaces are ubiquitous in the modern description  of amplitudes:
as string world--sheets describing interactions of  strings, in the Cachazo--He--Yuan (CHY) formalism and twisted intersection theory.
Formulating amplitudes in terms of functions on Riemann surfaces allocates the power of algebraic geometry to derive new amplitude relations between gauge and gravity and  enlarge our current understanding about symmetries and dualities in $D\!=\!4$. 
By holomorphic maps of a Riemann sphere into complex momentum space on a sphere  one can define the so--called scattering equations as  relating the space of kinematic invariants of $n$ massless particles and that of the positions of  $n$ points. These equations are the fundament of describing tree--level scattering of massless particles in any space--time dimensions as integrals over the moduli space of a punctured sphere \cite{CHYprl}. 
The CHY integrals are localized on the solutions of the scattering equations and yield the same propagators as in the field--theory limit of world--sheet string integrals \cite{Cachazo:2013iea}.
Indeed, the scattering equations have a natural appearance in string theory as saddle point equations 
describing its high energy limit \cite{Gross:1987ar,Gross:1987kza} and the sphere is the world--sheet describing the  tree--level  interactions
of closed strings. In string theory each $n$--point scattering process  is described by an underlying Riemann manifold with $n$ punctures accounting for the string world--sheet with $n$ vertex operator insertions. In fact, symmetries of the string world--sheet turn out to have profound impact on the structure of 
field--theory amplitudes itself and yield novel relations for field--theory. E.g.
the KLT relations follow from studying the analytic continuation  of string vertex operator positions on the Riemann sphere. Similarly, Kleiss--Kuijf and BCJ relations follow from considering world--sheet monodromies of open string vertex operators on the world--sheet disk  \cite{stie,Bjerrum-Bohr:2009ulz}. Similar results are  derived at one--loop \cite{Hohenegger:2017kqy,Tourkine:2016bak}, cf. also \cite{Casali:2019ihm,Stieberger:2021daa}.
In summary, both the CHY  formulae and string perturbation theory use a Riemann sphere to describe tree--level scattering and it is natural to ask how these approaches  are intertwined.

In fact, a supplementary   framework is provided by  intersection theory \cite{M2}. 
In this~setup tree--level amplitudes are  described by specifying a pair 
of  twisted differential forms on a Riemann sphere and their intersection numbers. 
 Twisted de Rahm cohomology  is suitable for considering multi--valued differential forms as they naturally appear on the string world--sheet. As a corollary  world--sheet string integrands can be recycled  for constructing twisted forms describing pure field--theory amplitudes. 
In this language the KLT relations can be interpreted as twisted period relations  \cite{Mizera:2017cqs}. Likewise, the CHY  formulae \cite{CHYprl} for pure gauge and gravitational amplitudes can be derived after specifying two differential forms and considering a special limit (corresponding to the high energy limit of string theory)  of their intersection numbers. More precisely, a saddle point approximation of the twisted period relations reproduces exactly the CHY formulae for pure both gauge and gravity amplitudes \cite{MT}.
It is one of the goals of this work to find the relevant differential forms describing EYM amplitudes and express the latter as intersection number of two twisted differential forms.
By all means the theory of twisted cycles and  intersection theory on Riemann surfaces nicely connects the formulae of the CHY formalism to string theory.

In addition, in the construction of ambitwistor string theories, which are chiral infinite tension analogues of ordinary RNS string theories, the amplitudes are localized on the scattering equations on genus zero Riemann surfaces \cite{Mason}.
The CHY formulae can be derived from a family of ambitwistor string theories and include universal representations  of different types of amplitudes such as pure gluon and pure graviton and the Einstein Yang--Mills scattering, cf. \cite{Casali:2015vta}.
These constructions rely on a pair of worldsheet systems providing world--sheet correlators,  which give rise to  the amplitudes. It has been shown that for pure gravity and pure Yang--Mills these result can be reproduced in  limits of twisted intersection numbers \cite{MT}. However, so far in this approach mixed amplitudes involving both gluons and gravitons could not be described in a straightforward way. In fact,  in here we shall suggest  an extension of disk correlators to sphere correlators by means of mapping open string fields to closed string fields. This is achieved by appending proper anti--holomorphic fields to the open string fields and 
adjusting the underlying conformal field theory.

In this work we shall provide all--multiplicity expressions for EYM amplitudes in terms of 
twisted intersection numbers. We shall proceed as follows. In section $2$ we  review some aspects of twisted intersection theory and the construction of certain field theory amplitudes in terms of intersections of twisted differential forms. We then demonstrate how the CHY formulae are recovered in the limit 
$\alpha'\!\ra\!\infty$. In section $3$ we present the necessary tools to compute disk amplitudes involving open and closed strings in terms of  Gra\ss mann variables. A detailed example for the simple case of two gluons and one graviton is provided.
We  calculate this superstring disk amplitude directly and write it in terms of fermionic variables. This gives an exact relation independent on $\ap$.
The comparison  with the corresponding CHY formula allows us to find the associated twisted differential forms. The generic expression of the twisted intersection numbers in the $\ap\!\ra\!\infty$ limit is then used to construct the EYM amplitude. 
In section $4$ we  propose an explicit embedding of the disk amplitude onto the sphere. Our embedding  is  compatible with the equations of motion of the string fields. We use this embedding to  construct the twisted differential forms for generic EYM amplitudes with an arbitrary number of gluons and gravitons. While the case of one graviton is devoted to section $5$ the case of an arbitrary number of gravitons is worked out in section $6$. 
We  first determine the full integrand of the underlying superstring disk amplitude in terms of fermionic variables, apply the embedding formalism and read off the candidates for a pair of twisted forms. 
Again,  the twisted intersection numbers of the latter reproduce in the $\ap\!\ra\!\infty$ limit the  EYM amplitudes. Furthermore, we present an all--multiplicity decomposition formula expressing any EYM amplitude  as linear combination over a minimal basis of pure  gluon subamplitudes.
Finally, in section 7 we give some concluding remarks and further directions. Appendix A is relegated to a unifying description of gluon and graviton world--sheet data on the  sphere and Appendix B contains some  supplementary material.

\sect{Amplitudes from Riemann surfaces with marked points}

In this section we present the necessary background for the present article. For a review and a detailed list of references  we refer the reader to \cite{MT}. 
We shall review some basics about both intersection theory and CHY formalism which both lead to a framework to define field--theories on a genus--zero Riemann surface $\Si$.
Despite  the field--theories are formulated on a string world--sheet in these setups not any  string $\ap$--corrections appear. We consider 
the configuration space $\mathcal{M}_{0,n}$ of $n\geq 3$ punctures $z_i\in \Sigma,\ i=1,\ldots,n$ on a genus--zero Riemann surface $\Si$.
The action of the automorphism group $SL(2,\IC)$ allows to fix
three points $z_j,z_k,z_l$ such that:
\begin{equation}\label{Space}
\mathcal{M}_{0,n}=\{(z_1,\ldots,z_i,\ldots,z_n) \in (\IC\IP)^{n-3}\ ,\ i\neq j,k,l \ |\  \mathop{\mathlarger{\forall}}_{m\neq n } z_m\neq z_n \}\ .
\end{equation}
The space \req{Space} gives rise to the moduli space   of Riemann spheres  with $n$ punctures.

\subsection{Amplitudes from twisted cohomology pairing}

Given a Riemann surface of genus zero   we introduce differential forms 
on the moduli space $\mathcal{M}_{0,n}$.
More precisely, due to the $SL(2,\IC)$ invariance we consider $(n-3)$--forms $\varphi$ as elements of cohomology equivalence classes
\begin{equation}\label{equivalence}
\varphi\simeq\varphi+\nabla_{\pm\omega} \xi\ ,
\end{equation}
with a rational $n-4$ form $\xi$ and the Gauss--Manin connection $\nabla_{\pm\omega}=d \pm \omega \wedge$ with $d$ the exterior derivative and some closed one--form (twist) $\omega$ to be specified later. Differential forms $\varphi$ with the property (\ref{equivalence}) give rise to classes $[\varphi]$ of twisted forms, which are $\nabla_{\pm\omega}$--closed modulo $\nabla_{\pm\omega}$--exact ones and belong to the 
$(n-3)$--th twisted cohomology group:
\begin{equation}\label{cohomology}
    H^{n-3}_{\pm\omega}(\mathcal{M}_{0,n},\nabla_{\pm\omega})=\frac{\{\varphi \in \Omega^{n-3}(\mathcal{M}_{0,n})\ |\ \nabla_{\pm\omega} \varphi=0\}}{\nabla_{\pm\omega} \Omega^{n-4}(\mathcal{M}_{0,n})}\ .
\end{equation}
The dual space $H^{n-3}_{-\omega}$ can be obtained from $H^{n-3}_{+\omega}$ by sending $\omega\ra-\omega$.
The intersection number on the twisted cohomology groups is the invariant pairing between  two forms 
$\varphi_\pm\in H^{n-3}_{\pm\omega}$ and defined by the integral
\begin{equation}
    \langle \varphi_+ , \varphi_- \rangle_\omega:=\lf(-\fc{\alpha'}{2\pi i}\ri)^{n-3}\ \int\limits_{\mathcal{M}_{0,n}} \iota_{\omega}(\varphi_+) \wedge \varphi_- \label{intersection}
\end{equation}
over  the space $\mathcal{M}_{0,n}$.
The map $\iota_{\omega}(\varphi_+)\in H^{n-3}_{+\omega}$ is the restriction of the twisted form over the compact support $H^{n-3}_{\omega,c}(\mathcal{M}_{0,n}, \nabla_{\omega})$  of $H^{n-3}_\omega(\mathcal{M}_{0,n},\nabla_{\omega})$. Otherwise, the integral over the moduli space $\mathcal{M}_{0,n}$ would not be well--defined since the latter is non--compact.
Examples of twisted forms are the $(n-3)$ (Parke--Taylor) forms
\be\label{PT}
PT(\si)=\fc{d\mu_n}{(z_{\si(1)}-z_{\si(2)})\ldots(z_{\si(n-1)}-z_{\si(n)})}\in H_{\pm\omega}^{n-3}\ \ \ ,\ \ \ \si\in S_n\ .
\ee
Above we have the measure
\begin{equation}\label{measure}
d\mu_n=z_{jk}z_{jl}z_{kl}\prod_{i=1\atop i\notin\{j,k,l\}}^ndz_i\ ,
\end{equation}
 which is a degree $n-3$ holomorphic form on $\mathcal{M}_{0,n}$, with $z_j,z_k,z_l$ being three arbitrary marked points fixed by $SL(2,\IC)$ invariance.

In order to make contact to physical amplitudes for the twist  $\omega$ one can select a potential $W$ with $\omega=dW$ as generating function  such that
\begin{equation}\label{potential}
    \omega=\alpha' \sum_{1\leq i, j \leq n } 2p_i p_j\ d\ln(z_i-z_j)\ ,
\end{equation}
with $n$ on--shell momenta $p_i$. The factor $\ap$ is chosen such that $\omega$ is dimensionless. One can see that the string Koba--Nielsen factor can be constructed in terms of $\omega$ as
\begin{equation}
 KN\equiv  \prod_{1\leq i, j \leq n} |z_i-z_j|^{ 2\alpha' p_i \cdot p_j}=e^{\int_{\gamma} \omega} \ ,\label{KN}
\end{equation}
for some path $\gamma$. Intersection numbers \req{intersection} are always rational functions of kinematic invariants with simple poles in the kinematic invariants.
A concrete computation of intersection numbers \req{intersection} yields \cite{MT}:
\be
\vev{PT(1,2,3,4),PT(1,2,4,3)}_\omega= \fc{1}{(p_1+p_2)^2}\ .
\ee
Here we evidence, that despite the definition  \req{intersection} and the potential \req{potential} involve higher orders in $\ap$ the 
localization procedure  entering the computation of the intersection number \req{intersection}
provides pure field--theory results. In fact, the intersection numbers \req{intersection}
localize near the boundary $\p\mathcal{M}_{0,n}$ of the moduli space where two or more points $z_i$  coalesce.

While the space \req{cohomology} of ordinary of $(n-3)$--forms  $\varphi$ is $(n-2)!$ dimensional the space of twisted $(n-3)$--forms is $(n-3)!$ dimensional, i.e. $\dim(H^{n-3}_\omega)=(n-3)!$.
Also the twisted homology group $H_{n-3}^\omega(\mathcal{M}_{0,n},KN)$, which is associated with the multivalued function $KN$, is $(n-3)!$ dimensional. Elements of the latter are specified by a cycle $C_a$ and local coefficients \req{KN}~as:
\be\label{twisthomo}
C_a\otimes KN\ .
\ee
The twisted homology cycles \req{cycles} are Poincare dual to the twisted cohomology $H_\om^{n-3}$ and one can consider the following pairing
\be\label{twistperiod}
\vev{C_a\otimes KN|\varphi_+}:=\int_{C_a}KN\ \varphi_+\ ,
\ee
which gives rise to period integrals on \req{Space}. 
Since $\dim H^{n-3}_\omega=\dim H_{n-3}^\omega=(n-3)!$ one can construct a basis
of $(n-3)!^2$ period integrals $\Pi^+_{ab}$ cf. \cite{Stieberger:2016xhs}.
A similar construction 
\be\label{dual1}
\tilde C_b\otimes KN^{-1}
\ee
applies for cycles $\tilde C_b$ of the  twisted homology group $H_{n-3}^{-\omega}(\mathcal{M}_{0,n},KN^{-1})$, which in turn gives rise to the period integrals:
\be\label{dual2}
\vev{\tilde C_b\otimes KN^{-1}|\varphi_-}:=\int_{\tilde C_b}KN^{-1}\ \varphi_-\ .
\ee
For two bases $\{C_a\}_{a=1}^{(n-3)!}$ and $\{\tilde C_b\}_{b=1}^{(n-3)!}$ of twisted cycles  we have the intersection matrix:
\begin{equation}
    \langle  C_a \otimes KN \;|\; \tilde C_b \otimes KN^{-1} \rangle=S_{ab}^{-1}\ .
\end{equation}
With these preparations we can  expand the intersection form  (\ref{intersection})   as
\begin{align}
    \langle \varphi_+,\varphi_- \rangle_\om&=
    \sum_{a,b=1}^{(n-3)!} \vev{C_a\otimes KN|\varphi_+} \   \langle  C_a \otimes KN | \tilde C_b \otimes KN^{-1} \rangle\   \vev{\tilde C_b\otimes KN^{-1}|\varphi_-}\nonumber\\
    &=\lf(\fc{\alpha'}{2\pi i}\ri)^{n-3}\ \sum_{a,b=1}^{(n-3)!} \lf(\int_{C_a} KN\ \varphi_+\ri)\  S_{ab}^{-1} \ \lf(\int_{\tilde C_b} KN^{-1}\ \varphi_-\ri)\ ,\label{TRPR}
\end{align}
which is equivalent to the twisted Riemann's period relations by Cho and Matumoto \cite{Cho}.

There is an isomorphism between the dual twisted cohomologies thanks to Hanamura and Yoshida \cite{Hanamura}: 
\be\label{ISO}
H^{n-3}_{-\omega}\simeq H^{n-3}_{\overline\omega}\ .
\ee
Then, the dual homology objects \req{dual1} and \req{dual2} can be formulated  by replacing $KN^{-1}$ and $\overline{KN}$.
The pairing $H^{n-3}_{\omega}$ and $H^{n-3}_{\overline\omega}$ is more suited to describe closed string world--sheet integrals. In particular, in this language  the twisted period relations \req{TRPR} can be interpreted as KLT relations  \cite{Mizera:2017cqs}. 
We shall make use of this isomorphism when constructing our twisted forms for EYM.

To make contact with amplitudes let us present examples of twisted forms. 
World--sheet string correlators are borrowed to construct 
these twisted forms for field--theory. Plugging the latter into the intersection number \req{intersection} yields  field--theory amplitudes.

\subsubsection{Twisted form for the color sector} 

\noindent
The color ordered twisted form is given by \req{PT} \cite{M1}:
\begin{equation}\label{Color}
    \varphi^{color}_{n}=d\mu_{n}\ \frac{\Tr(T^{c_1}T^{c_2}\ldots T^{c_n})}{(z_1-z_2)(z_2-z_3)\ldots (z_n-z_1)}\equiv \Tr(T^{c_1}T^{c_2}\ldots T^{c_n})\ PT(1,2,\ldots,n)\ .
\end{equation}
The generators $T^c$ carry the color degrees of freedom of a non--Abelian gauge group.
Plugging \req{Color} and permutations thereof into  (\ref{intersection}) provides the double ordered subamplitudes of the bi--adjoint scalar theory  (cf. \cite{M1}). E.g. we have:
\begin{equation} \langle \varphi^{color}_{4},\varphi^{color}_{4}\rangle_\om=
\Tr(T^{c_1}T^{c_2}T^{c_3} T^{c_4})\Tr(T^{c_1}T^{c_2}T^{c_3}T^{c_4})\ \lf\{\fc{1}{(p_1+p_2)^2}
+\fc{1}{(p_1+p_3)^2}\ri\}\ .
\label{colorinter}
\end{equation}

\subsubsection{Twisted form for the gauge and gravity sector} 
\label{Twistedgaugegravity}

\noindent
For the twisted form describing gauge interactions one can use an expression familiar from superstring theory  (cf. \cite{M1}) 
\begin{equation}\label{Witten}
\begin{aligned}
\varphi^{gauge}_{\pm,n}=d\mu_{n} \int \prod\limits_{i=1}^{n} d \theta_i d\bar{\theta_i}\frac{\theta_k \theta_l}{z_k -z_l} \exp\Bigg\{-\sum\limits_{i\neq j}\frac{ \theta_i \theta_j p_i \cdot p_j+\bar{\theta}_i \bar{\theta}_j\varepsilon_i \cdot \varepsilon_j +2(\theta_i - \theta_j)\bar{\theta_i}\varepsilon_i \cdot p_j}{z_i-z_j \mp \alpha'^{-1} \theta_i \theta_j}\Bigg\}\ ,
\end{aligned}
\end{equation}
with the gauge boson polarization vectors $\eps^\mu_i,\ i=1,\ldots,n$. The twisted form
\req{Witten} involves an integral over $2n$ Gra\ss mann variables $\theta_i,\bar\theta_i$. 

The two twisted forms \req{Color} and \req{Witten} can be used to compute
the YM partial subamplitude from the  intersection number \req{intersection}:
\begin{equation}
\begin{aligned}
\vev{\varphi^{color}_{n},\varphi^{gauge}_{+,n}}_\omega=\mathcal{A}_{YM}(1,2,\ldots,n)\ \Tr(T^{c_1}T^{c_2}\ldots T^{c_n})\ .
\end{aligned}\label{gaugeinter}
\end{equation}

Furthermore, the twisted forms  \req{Witten} can be used to compute
the gravitational amplitude from the  intersection \req{intersection}:
\begin{equation}
\begin{aligned}
\vev{\varphi^{gauge}_{-,n},\varphi^{gauge}_{+,n}}_\omega=\mathcal{M}(1,2,\ldots,n)\ .
\end{aligned}\label{gravityinter}
\end{equation}
Note that after computing the intersection number \req{intersection} in the results 
\req{colorinter}, \req{gaugeinter} and \req{gravityinter} the inverse string tension $\ap$ drops out and yields a pure field--theory result. This property generally applies for intersection numbers of logarithmic forms, which have at most simple poles along the boundary divisor 
$\partial \mathcal{M}_{0,n}$ \cite{M2}.

\subsubsection{Relation to string amplitudes}
\noindent
A possible choice of homology basis \req{twisthomo} describes integration cycles   $C_a$  for color ordered open string subamplitudes (corresponding to the three fixed points $z_1,z_{n-1},z_n$):
\be\label{cycles}
C_a=\{(z_2,\ldots,z_{n-2})\in \IR^{n-3}\ |\ z_{a(2)}<\ldots<z_{a(n-2)}\}\ ,\ a\in S_{n-3}\ .
\ee
Together with the  twisted gauge form \req{Witten} the $n$--point open superstring subamplitude $\Ac^{string}_n(a)$ corresponding to the color ordering $a$ can be written as twisted period \req{twistperiod}:
\be
\Ac^{string}_n(a)=\int_{C_a} KN\ \varphi^{gauge}_{+,n}\ .
\ee
In the low energy limit it reproduces the field--theory gauge subamplitude \req{gaugeinter}, i.e.:
\begin{equation}
\lim\limits_{\alpha' \rightarrow 0}\ \int\limits_{C_a} KN \ \varphi^{gauge}_{+,n}=\vev{ PT(a), \varphi^{gauge}_{+,n}}_\om\ .
\end{equation}
The period matrix $F_{ab}\equiv \Pi^+_{ab}$ encoding the period integrals \req{twistperiod}   specifies  the full open superstring 
amplitude  \cite{Mafra:2011nv,Mafra:2011nw,Schlotterer:2012ny,Broedel:2013tta,Stieberger:2016xhs}.

\subsection{Scattering equations from intersection of twisted forms and $\ap\ra\infty$ limit} 

The one--form \req{potential} is related to the scattering equations in the limit $\ap\ra\infty$ describing the  massless limit.
In fact, in this limit we have $\nabla_{\pm\omega}\ra \pm \omega \wedge$. 
Then the space \req{cohomology} becomes the single space of $(n-3)$--forms subject to additions of $\omega=0$, which gives rise to the scattering equations:  
\begin{equation}\label{SEQ}
f_k:=\sum_{j\neq k}\frac{p_k p_j}{z_j-z_k}=0\ ,\ \ 1\leq k\leq n\ .
\end{equation}
We consider the relation \req{TRPR} for a basis of orthonormal cycles $S_{ab}=\delta_{ab}$, which for \req{KN}  can be written as
\be\label{trpr}
 \langle \varphi_+,\varphi_- \rangle_\omega=\lf(\fc{\alpha'}{2\pi i}\ri)^{n-3}\  \sum_{a=1}^{(n-3)!} \lf(\int_{C_a} e^{\int_{\gamma} \omega} \; \varphi_+\ri) \lf( \int_{\tilde C_a} e^{-\int_{\gamma} \omega} \; \varphi_-\ri) \ ,
 \ee
for some path $\gamma$. For $\alpha' \rightarrow \infty$ with  $\varphi_\pm=\hat{\varphi}_\pm \,\,d^n z$ the saddle point approximation limit 
 of \req{trpr} yields 
\begin{align}
         \lim_{\alpha' \rightarrow  \infty}\langle \varphi_+,\varphi_- \rangle_\omega&= \lf(\fc{\alpha'}{2\pi i}\ri)^{n-3} \lf. \sum_{a=1}^{(n-3)!}  
         \lf( \frac{(2\pi)^{\h(n-3)}}{\sqrt{\det(-\frac{\partial^2 \int \omega}{\partial z_i \partial z_j})}}\; \lim_{\alpha' \rightarrow  \infty} \hat{\varphi}_+\ri)\!\!\! \lf(\frac{(2\pi)^{\h(n-3)}}{\sqrt{\det(\frac{\partial^2 \int \omega}{\partial z_i \partial z_j})}} \;  \lim_{\alpha' \rightarrow  \infty} \hat{\varphi}_-\ri)\!\!\ri|_{z_i=z_i^{(a)}}\hskip-1cm \nonumber\\
         & =(-\alpha')^{n-3} \sum_{a=1}^{(n-3)!} \Bigg( \det \frac{\partial^2 \int \omega}{\partial z_i \partial z_j} \Bigg) ^{-1} \lf.\lim_{\alpha' \rightarrow  \infty} \hat{\varphi}_+ \;\hat{\varphi}_-\ri|_{z_i=z_i^{(a)}}\ ,
    \end{align}
with the $(n-3)!$ saddle points $(z_2^{(a)},\ldots,z_{n-2}^{(a)}),\; a=1,\ldots,(n-3)!$.
In particular, for the case at hand \req{potential} we arrive at
\be
 \lim_{\alpha' \rightarrow  \infty}\langle \varphi_+,\varphi_- \rangle_\omega=
\alpha'^{\frac{n-3}{2}} 
    \int\limits_{\mathcal{M}_{0,n}} d\mu_n\ \prod_{k=2}^{n-2} \delta(f_k) \  \lim_{\alpha' \rightarrow  \infty} \hat{\varphi}_+\; \hat{\varphi}_-\ ,
    \label{expansion}
\end{equation}
which yields the localization over the scattering equations \req{SEQ} of the CHY formulae. 
As a consequence in the limit $\ap\ra\infty$ intersection numbers become scattering amplitudes formulated in the CHY formalism. 
Note, that for the three examples \req{colorinter}, \req{gaugeinter} and \req{gravityinter}
the relation \req{expansion} is exact and independent on $\ap$. This is generically the case for logarithmic twisted forms \cite{M2}.
On the other hand, as we shall see in this work this is not obvious for EYM amplitudes where 
the limit $\ap\ra\infty$ will apparently also give rise to subleading terms in $\ap^{-1}$.

%%%%%%%%%%%%%%%%%%%%%%%%%%%%%%%%%%%%%%%%
\subsection{CHY formalism}

The CHY formalism provides formulae for scattering amplitudes as a sum over solutions of the scattering equations \req{SEQ} of certain Pfaffian and determinant expressions 
\cite{CHYprl,Cachazo:2013iea,chy}.
One of the main results of the present work will be to construct a pair of twisted forms  
$\varphi^{EYM}_\pm,\widetilde\varphi^{EYM}_\pm$ which after applying in \req{intersection} leads to the CHY formula for EYM amplitudes in the large $\ap$--limit \req{expansion}.
Therefore, here we present a brief introduction to the CHY formalism and its notations which we shall use later.

In the CHY formalism the generic EYM amplitude of $n$ gluons and $r$ gravitons is formulated as
integral over \req{Space} localizing at the solutions of the scattering equations \req{SEQ} \cite{chy}:
\begin{equation}\label{AmpEYM}
\mathcal{A}_{CHY}(n;r)=\int\limits_{\mathcal{M}_{0,n+r}} d \mu_{n+r}\ \sideset{}{'}\prod_{a=1}^{n+r} \delta(f_a) \ \mathcal{I}_{n+r}(p,\varepsilon,\sigma)  \ .
\end{equation}
Note, that this amplitude refers to a single trace color structure of the form $\Tr(T^{c_1}\ldots T^{c_n})$. Above, the localization of the amplitude at the scattering equation \req{SEQ} is achieved through the delta--functions $\delta(f_a)$ with support on
\begin{equation}
f_a \equiv \sum\limits_{b=1\atop b\neq a}\frac{p_a  p_b}{\sigma_a-\sigma_b} \ \ \ ,\ \ \ a=1,\ldots,n+r\ ,
\label{f}
\end{equation} 
and  $p_a$ denote the set of both gluon and graviton momenta, which we shall properly introduce in \ref{Notation}. In \req{AmpEYM} the prime at the product means that due to $SL(2,\IC)$ invariance three of the delta functions must be removed and the measure $d\mu_{n+r}$ refers to the gauge fixed expression \req{measure} w.r.t.  $n+r$ coordinates $\sigma^a$.
Finally, the integrand $\mathcal{I}_{n+r}$ is defined as
\begin{equation}
    \mathcal{I}_{n+r}(1,2,\ldots,n;1,2,\ldots,r)=\mathcal{C}(1,2,3,...,n)\ \pf \Psi_{S_r}\ \pf'\Psi_{S}(k_a,q_a,\varepsilon,\sigma)\ , \label{gchy}
\end{equation}
with gluon momenta $k_a$  and graviton momenta  $q_a$ (all collectively denoted by $p_l$).
The set  $S_r\!=\!\{i_1,\ldots,i_r\}$ encompasses all $r$ graviton labels $i_l$, while the set $S\!=\!\{i_1,\ldots,i_{n+r}\}$ packages all 
$n\!+\!r$ gluon and graviton numbers. If not otherwise stated we have 
$S_r\!=\!\{n+1,\ldots,n+r\}$ and write $\Psi_{S_r}=\Psi_r$ and $\Psi_{S}=\Psi_{n+r}$. Here, $\pf'$ is the reduced Pfaffian
\begin{equation}\label{redPF}
\pf'\Psi \equiv \frac{(-1)^{i+j}}{\sigma_i-\sigma_j}\pf (\Psi)^{ij}_{ij}\ ,
\end{equation}
with the index $i,j$ denoting removals of  rows $i,j$ and columns $i,j$.

For the specific case of one graviton the integrand \req{gchy} can be written as
\begin{equation}
    \mathcal{I}_{n+1}(1,2,...,n;q)=\mathcal{C}(1,2,...,n)\ C_{qq}\ \pf'\Psi_{n+1}(k_a,q,\varepsilon,\sigma)\ , \label{chy}
\end{equation}
with $\mathcal{C}(1,2,...,n)$  the Parke--Taylor (cf. also \req{PT}) factor given by:
\begin{equation}\label{ColorForm}
\mathcal{C}(1,2,...,n)=\frac{1}{(\sigma_1-\sigma_2)(\sigma_2-\sigma_3)...(\sigma_{n}-\sigma_1) }\ .
\end{equation}
Furthermore, for  $C_{qq}$ we have (with $\si_{l,m}=\si_l-\si_m$)
\begin{equation}
C_{qq}=\sum\limits_{l=1}^{n-1} (\varepsilon_q \cdot  x_l)\ \frac{\sigma_{l,l+1}}{\sigma_{l,q}\sigma_{l+1,q}}= \pf \Psi_{S_1}\equiv \pf \Psi_1\ , \label{cqq}
\end{equation}
with the definition  $x^\mu_l=\sum\limits_{j=1}^{l} p^\mu_j$ and $S_1\!\equiv\!\{q\}$ with $q\!\equiv\! n+1$. 
For a given subset  $S=\{i_1,\ldots,i_m\}$ the kinematic matrix $\Psi_S$ is defined as  $2m\times 2m$--matrix\footnote{In some references the submatrices $A$,$B$ and $C$ are multiplied by a factor of two \cite{chy}. Here we shall use the convention of \cite{MT,p}.} 
\be\Psi_{S}=  \begin{pmatrix} A & -C^T \\ C & B \end{pmatrix}\ ,
 \label{PSI}
 \ee
with the three $r\times r$--submatrices:
\be  A= \begin{cases}  0\ , \,\,\,\,\,\,\,\, i=j \\
                    \frac{p_i p_j}{\sigma_i-\sigma_j}\ , \,\,\,\,\,\,\,\, i \neq j  
       \end{cases},\  \,\,
 C = \begin{cases}  -\sum\limits_{k \neq i}^{n+r}\frac{\varepsilon_i p_k}{\sigma_i-\sigma_k}\ , \,\,\,\,\,\,\,\, i=j \\
                   \frac{\varepsilon_i p_j}{\sigma_i-\sigma_j}\ ,  \,\,\,\,\,\,\,\,i \neq j 
       \end{cases}, \,\,
  B= \begin{cases}  0\ , \,\,\,\,\,\,\,\, i=j\ , \\
                    \frac{\varepsilon_i  \varepsilon_j}{\sigma_i-\sigma_j}\ ,\,\,\,\,\,\,\,\, i \neq j\ .
       \end{cases}\label{psi2}
 \ee

\sect{From disk amplitudes to twisted forms for Einstein--Yang Mills}
\label{Section3}

The object \req{Witten} originates in string theory  to compactly write superstring amplitudes \cite{witt}. 
String amplitudes are described by a  punctured Riemann surface with each puncture representing a vertex operator position associated to creation or annihilation of a string state.  E.g. the closed superstring tree--level amplitude assumes the form
\begin{equation}
   \begin{aligned}
    & \mathcal{A}(n)=\int\limits_{\mathcal{M}_{0,n}}  d\mu_n\ \vev{ V(\varepsilon_1,p_1,z_1)...V(\varepsilon_n,p_n,z_n)}=\int\limits_{\mathcal{M}_{0,n}}  \prod\limits_{i<j}^{n} |z_i-z_j|^{2\alpha' p_i \cdot p_j}\  |F (p,\varepsilon)|^2\ ,
\end{aligned} \label{wickamp}
\end{equation}
with the measure \req{measure} and  the twisted gauge form \req{Witten}:
\begin{equation}
\begin{aligned}
&& F(p,\varepsilon)=d\mu_{n} \int \prod\limits_{i=1}^{n} d \theta_i d\bar{\theta_i}\ \frac{\theta_k \theta_l}{z_k -z_l} \exp\Bigg(\sum\limits_{i\neq j}\frac{ \theta_i \theta_j p_i \cdot p_j+\bar{\theta}_i \bar{\theta}_j\varepsilon_i \cdot \varepsilon_j +2(\theta_i - \theta_j)\bar{\theta_i}\varepsilon_i \cdot p_j}{z_i-z_j - \alpha'^{-1} \theta_i \theta_j}\Bigg)\ .
\end{aligned}\label{twistedWitten}
\end{equation}

In this work we deal with the massless states of the  closed  and open superstring
describing a graviton and a gluon, respectively. Scattering amplitudes of open and closed
strings at tree--level are described by a disk world--sheet.
In the sequel we shall introduce some necessary tools for computing disk amplitudes. Later  we shall use the latter to construct twisted forms for EYM amplitudes.
For this it is instrumental  that we look at the standard vertex operators of superstring theory in terms of Gra\ss mann variables  \cite{witt}. 

\subsection{Vertex operators and disk correlator}
\label{vertexO}

 On the world--sheet disk open string vertex positions $x_i$ are located at the boundary of the disk, while closed string vertex positions $z_i$ are inserted in the bulk of the disk. The momentum  and polarization of a massless gluon are given by $k_\rho$ and $\eps_\mu$ while those of a massless graviton by   $q_\rho$
and $\eps_{\mu\nu}$, respectively.
To cancel the ghost background charge on the genus zero Riemann surface  we  define their vertex operators in the zero and $(-1)$--ghost pictures. We have the following list of  vertex operators:
\begin{itemize}
\item  Open string vertex operator  in zero--ghost picture:
\begin{align}
 V^{(0)}_o(x,\varepsilon,k)&=\varepsilon_\mu\ [\partial X^\mu+\alpha'(k\cdot \psi) \psi^{\mu}]\ e^{ik \cdot X}\nonumber\\
 &=\int d\theta d\bar{\theta}\ \exp\lf\{ik\cdot X+\theta\bar{\theta} \varepsilon \cdot \partial X+\theta \sqrt{\alpha
'} k\cdot \psi+\bar{\theta}\sqrt{\alpha
'} \varepsilon \cdot \psi\ri\} \label{Open0}
\end{align}
 \item Closed string vertex operator in $(0,0)$--ghost picture: \begin{align} 
  V_c^{(0,0)}(z,\ov{z},\varepsilon,q) & =\varepsilon_{\mu \nu} \, \Big[ i\ov{\partial}\widetilde{X}^\mu +\frac{\alpha'}{2} (\widetilde{q}\widetilde{\psi})\widetilde{\psi}^\mu(\ov{z}) \Big] \Big[ i \partial X^\nu +\frac{\alpha'}{2} (q \psi) \psi^\nu(z) \Big]  \, e^{iqX(z, \ov{z})}\nonumber\\
    &=\int d\theta_1 d\bar{\theta}_1 d\theta_2 d\bar{\theta}_2\ \exp\Big\{i q \cdot X+i \widetilde{q}\cdot \widetilde{X}+\theta_1\bar{\theta}_1 \varepsilon \cdot \partial X+\theta_1 \sqrt{\alpha
'} q\cdot \psi\nonumber\\
    & +\bar{\theta}_1 \sqrt{\alpha
'} \varepsilon \cdot \psi+\theta_2\bar{\theta}_2 \varepsilon \cdot \ov{\partial}\widetilde{X}+\theta_2 \sqrt{\alpha
'} \widetilde{q}\cdot \widetilde{\psi}+\bar{\theta}_2 \sqrt{\alpha
'} \varepsilon \cdot \widetilde{\psi}  \Big\}\ .
\label{Grav1}\end{align}
   
 \item Open string vertex operator  in the $(-1)$--ghost picture:
 \be  V_o^{(-1)}(x,\alpha,k)=\varepsilon_\mu\ e^{-\phi(x)}\psi^\mu(x) \ e^{i k\cdot X(x)}=\int d\theta \exp\lf\{ik\cdot X+\theta \varepsilon \cdot \psi\ri\}\ .\label{Open1}
 \ee
     \item Closed string vertex operator in the $(-1,-1)$ picture:
     \begin{align}
    V_c^{(-1,-1)}(z,\ov{z},\varepsilon,q)&=\varepsilon_{\mu \nu}\ e^{-\bar{\phi}(\bar{z})} \widetilde{\psi}^\mu(\bar{z}) \ e^{-\phi(z)}  \psi^\nu (z) \ e^{iq \cdot X(z,\bar{z})}\nonumber\\
     & =\int d\theta d\bar{\theta}\ \exp\lf\{i q \cdot X+i \widetilde{q}\cdot \widetilde{X}+\bar{\theta} \widetilde{\varepsilon}\cdot \widetilde{\psi}+\theta \varepsilon \cdot \psi  \ri\}\ .
   \label{Grav2}\end{align}
\end{itemize}
Finally, for the string $S$--matrix the following on--shell conditions must be imposed: 
\begin{equation}
\begin{aligned}
 & k^2=q^2=0\ , \\
 & k^\mu  \varepsilon_\mu=0\ , \qquad q^\mu \cdot \varepsilon_{\mu\nu}=0\ , \qquad \varepsilon^\mu_\mu =0\ .
\end{aligned} \label{onshell}
\end{equation}
These are the standard massless, transverse and traceless conditions of on--shell string states. 

We shall be interested in the  disk amplitudes\footnote{The open string vertices \req{Open0} and \req{Open1} carry also gauge degrees of freedom
accounted for by Chan--Paton factors $T^c$. Their respective ordering in the disk amplitude 
provides the color ordering of the amplitude. Since in the sequel we shall only be concerned with the integrand of the amplitude we can drop all Chan--Paton factors.} $\mathcal{A}(n;r)$ involving $n$ open strings  and $r$ closed strings. This leads to the following disk correlator
\be
  \Big\langle \prod\limits_{i=1}^{n} V_o(\varepsilon_i,k_i,x_i) \prod\limits_{s=1}^{r} V_c (\varepsilon_s,q_s,z_{n+s},\bar{z}_{n+s})\Big\rangle_{D^2}\ ,
     \label{genz}
\ee
involving the above vertex operators. On the disk boundary conditions have to be imposed
for the closed string fields. In the following we shall only be concerned with Neumann
boundary conditions leading to $q_s=\tilde q_s$.
Then, the  correlator can be evaluated by standard conformal  field theory techniques. At this point it is important to notice that 
there is a universal Koba--Nielsen factor \cite{stie}
\begin{align}
KN_{D_2}&=\prod_{i<j}^n |x_j-x_i|^{2\ap k_ik_j}\ \prod_{a<b}^{r}|z_{n+b}-z_{n+a}|^{\ap q_aq_b}
|z_{n+b}-\ov z_{n+a}|^{\ap q_aq_b}\nonumber\\ 
&\times\prod_{i=1}^n\prod_{a=1}^{r} |z_{n+a}-x_i|^{2\ap k_iq_a}\ ,\label{KND2}
\end{align}
involving open string positions $x_i$ and closed string positions $z_{n+a}$ and external momenta
$k_i$ and $q_a$. 
We relegate a universal description  for the  momenta, polarizations and positions of the string vertex operators into the  \ref{Notation}.

\subsection{Amplitude of two gluons and one graviton}

As a first step we shall look at the simple example of an amplitude involving two gluons and one graviton represented in string theory by a world--sheet disk with two open strings and one closed string. Computing the latter and putting it into the form  \req{twistedWitten} will allow us to extract the twisted forms describing the EYM amplitude.

\subsubsection{Disk amplitude of two gluons and one graviton}

In this subsection we compute the string amplitude involving two gluons and one graviton.
The corresponding amplitude is described by a disk world--sheet   with two open and one closed string state. We  take the two open string vertex operators in the $(-1)$--ghost picture to furnish the required total background  charge of $(-2)$.
Using the standard notion of correlators and vertex operators  \req{Grav1} and \req{Open1}    we write the amplitude as:
\begin{align}
      \mathcal{A}(2;1)&=\int \frac{dz_1dz_2d^2z_3}{\textit{SL}(2,\IR)} \ 
      \langle V_o^{(-1)}(\varepsilon_1,k_1,z_1) V_o^{(-1)}(\varepsilon_2,k_2,z_2)V^{(0,0)}_c(\varepsilon_q,q,z_3,\bar{z}_3)\rangle\nonumber \\
    &=\varepsilon_{\mu}\varepsilon_{\nu} \varepsilon_{\alpha\beta}\ \int \frac{dz_1dz_2d^2z_3}{\textit{SL}(2,\IR)}\ \Bigg\langle   e^{-\phi(z_1)}\psi^{\mu} (z_1) e^{i k_1\cdot X(z_1)}\   e^{-\phi(z_2)}\psi^{\nu} (z_2) e^{i k_2\cdot X(z_2)}\label{AMPEYM21}\\
   & \hskip2cm  \times\Big[ i\ov{\partial}\widetilde{X}^\alpha +\frac{\alpha'}{2} (q\widetilde{\psi})\widetilde{\psi}^\alpha(\ov{z}_3) \Big] \Big[ i \partial X^\beta +\frac{\alpha'}{2} (q \psi) \psi^\beta(z_3) \Big]  \, e^{iqX(z_3, \ov{z}_3)}\, \Bigg\rangle_{D^2}\ .\nonumber
   \end{align}
   The open string vertex positions $z_1,z_2$ are integrated along the boundary of the disk, while the closed string position $z_3$ is integrated over the full disk respecting the $SL(2,\IR)$ symmetry of the disk world--sheet. 
The amplitude \req{AMPEYM21} has a particular kinematic structure. We have the following 
on--shell constraints \req{onshell} for the polarization and momenta of the states:
\begin{equation}
\begin{aligned}
     \varepsilon_{\alpha\beta}&=\varepsilon_{\alpha} \otimes \varepsilon_{\beta}\ ,\\
   k^{\mu} \varepsilon_\mu&=0,\,\,\, q^\alpha \varepsilon_{\alpha\beta}=0\ , \\
   k_1 \cdot k_2&=k_1\cdot q =k_2\cdot q =0\ , \\
  k_1 \cdot k_2& =k_1\cdot \widetilde{q} =k_2\cdot \widetilde{q} =0\ . \label{shell}
  \end{aligned}
\end{equation}
Due to the on--shell  massless  three--particle conditions \req{shell} the Koba--Nielsen factor $KN_{D_2}$ stemming from the exponentials \req{KND2} becomes trivial:
\be\label{KNtrivi}
KN_{D_2}=|z_1-z_3|^{2\alpha' p_1 \cdot q}|z_2-z_3|^{2\alpha' p_2 \cdot q}|z_1-z_2|^{2\alpha' p_1 \cdot p_2}=1\ .
\ee
After performing in \req{AMPEYM21} the remaining contractions we get
  \begin{equation}
    \begin{aligned}
   \mathcal{A}(2;1)  & =C \int \frac{dz_1dz_2d^2z_3}{\textit{SL}(2,\IR)}  \  \frac{\varepsilon_\mu \varepsilon_\nu \varepsilon_\alpha \varepsilon_\beta}{z_1-z_2} \\
    & \times \Bigg\{\frac{g^{\mu\nu}}{z_1-z_2}\Bigg( \frac{g^{\alpha\beta}}{(z_3-\bar{z}_3)^2}+\frac{k_1^{\alpha}z_{12}}{(z_1-\bar{z}_3)(z_2-\bar{z}_3)} \frac{k_2^{\beta}z_{12}}{(z_1-z_3)(z_2-z_3)} \Bigg)\\
    &\ \  +\frac{1}{2}\ \frac{k_1^{\alpha}z_{12}}{(z_1-\bar{z}_3)(z_2-\bar{z}_3)}\Bigg(-\frac{q^\mu}{z_1-z_3}\frac{g^{\nu\beta}}{z_2-z_3} +\frac{q^\nu}{z_1-z_3}\frac{g^{\mu\beta}}{z_2-z_3}\Bigg)\\
    &\ \  +\frac{1}{2}\ \frac{k_1^{\beta}z_{12}}{(z_1-z_3)(z_2-z_3)}\Bigg(-\frac{q^\mu}{z_1-\bar{z}_3}\frac{g^{\nu\alpha}}{z_2-\bar{z}_3} +\frac{q^\nu}{z_1-\bar{z}_3}\frac{g^{\mu\alpha}}{z_2-\bar{z}_3}\Bigg) \Bigg\} \ ,   \end{aligned} \label{1}
\end{equation}
which can be simplified to
{\small \begin{align} 
\mathcal{A}(2;1)&=   C \int \frac{dz_1dz_2d^2z_3}{\textit{SL}(2,\IR)}\ \Bigg\{\frac{(\varepsilon_3  k_1)}{(z_1-z_3)(z_1-\bar{z}_3)(z_2-z_3)(z_2-\bar{z}_3)}\Bigg[-\frac{1}{2}(\varepsilon_2  \varepsilon_4)(\varepsilon_1  q)+\frac{1}{2}(\varepsilon_1 \varepsilon_4)(\varepsilon_2  q)\Bigg] \nonumber\\[2mm]
    & +\frac{(\varepsilon_4  k_1)}{(z_1-z_3)(z_1-\bar{z}_3)(z_2-z_3)(z_2-\bar{z}_3)}\Bigg[-\frac{1}{2}(\varepsilon_2  \varepsilon_3)(\varepsilon_1  q)+\frac{1}{2}(\varepsilon_1  \varepsilon_3)(\varepsilon_2 \ q)+(\varepsilon_1  \varepsilon_2)(\varepsilon_3 \cdot k_2)\Bigg]\Bigg\},\label{expression}
      \end{align}}
with some normalization constant $C$. Note, that the expression \req{expression} is symmetric under the exchange $\eps_3\leftrightarrow \eps_4,\ z_3\leftrightarrow \bar z_3$. Imposing the standard double copy structure for the graviton polarization as $\varepsilon_4=\widetilde{\varepsilon}_3=\varepsilon_3$ we have:
     \begin{equation}
    \begin{aligned}
     \mathcal{A}(2;1) &= C\int \frac{dz_1dz_2d^2z_3}{\textit{SL}(2,\IR)}  \  \frac{(\widetilde{\varepsilon}_{3}  k_1)}{|(z_1-z_3)|^2|(z_2-z_3)|^2}\\
     &\times\Bigg\{-(\varepsilon_2  \varepsilon_3)(\varepsilon_1  q)+(\varepsilon_1 \varepsilon_3)(\varepsilon_2  q)+(\varepsilon_1  \varepsilon_2)(\varepsilon_3 k_2)\Bigg\}\ .
    \end{aligned} \label{2a}
\end{equation}
Note, that the result \req{2a} is exact to all orders in $\ap$. The calculations we have done so far are standard string amplitude computations.
In the following we shall only be interested in the integrand of \req{2a} and treat the latter as complex function depending on the three  vertex positions $z_i\in \IC$. 
Following the steps after \req{wickamp} we shall now introduce fermionic variables $\theta_i,\bar\theta_i$ to describe the integrand of the amplitude \req{AMPEYM21}.
We can write the integrand of \req{2a} as
\begin{equation}
    \begin{aligned}
  & \mathcal{I}(2;1) =\Bigg\langle \int d\theta_1 d\ov{\theta}_1  d\theta_2 d\bar{\theta}_2 d\theta_3 d\bar{\theta}_3 d\theta_4 d\bar{\theta}_4\frac{\theta_1 \theta_2}{z_1-z_2}\\
  & \hspace{2cm}\times\exp[ik_1\cdot X+\theta_1\bar{\theta}_1 \varepsilon_1 \cdot \partial X+\theta_1 \sqrt{\alpha'} k_1\cdot \psi+\bar{\theta}_1\sqrt{\alpha'}  \varepsilon_1 \cdot \psi] \\
  &  \hspace{2cm}\times \exp[ik_2\cdot X+\theta_2\bar{\theta}_2 \varepsilon_2 \cdot \partial X+\theta_2 \sqrt{\alpha'} k_2\cdot \psi+\bar{\theta}_2 \sqrt{\alpha'}  \varepsilon_2 \cdot \psi]\\
    &  \hspace{2cm}\times  \exp[i q \cdot X+i \widetilde{q}\cdot \widetilde{X}+\theta_3\bar{\theta}_3 \varepsilon_3 \cdot \partial X+\theta_3 \sqrt{\alpha'} q\cdot \psi+\bar{\theta}_3 \sqrt{\alpha'} \varepsilon_3 \cdot \psi\\
    &   \hspace{2.6cm} +\theta_4\bar{\theta}_4 \widetilde{\varepsilon}_{3} \cdot \ov{\partial}\widetilde{ X} +\theta_4 \sqrt{\alpha'} \widetilde{q}\cdot \widetilde{\psi}+\bar{\theta}_4 \sqrt{\alpha'}\widetilde{\varepsilon}_{3} \cdot \widetilde{\psi}  ] \Bigg\rangle_{D^2}\ ,
   \end{aligned}\label{step1}
\end{equation} 
which in turn can be expressed  in terms of the Gra\ss mann integral  \req{twistedWitten}. We open up the sums above and expand the exponential up to order in the fermionic  variables $\theta_i,\bar\theta_i$ which leads to a  non-vanishing Gra\ss mann integral:
\begin{align}
    \mathcal{I}(2;1)  &=\ap^2 \int \prod\limits_{i=1}^{4}  \frac{\theta_1 \theta_2}{z_1-z_2} d \theta_i d\ov{\theta}_i \Bigg\{\frac{(\bar{\theta}_1 \theta_3 \varepsilon_1 \cdot q)  (\ov{\theta}_3 \ov{\theta}_2 \varepsilon_3 \cdot \varepsilon_2)}{(z_1-z_3)(z_3-z_2)}\theta_4 \ov{\theta}_4\Bigg( \frac{(\widetilde{\varepsilon}_{3}  \cdot k_1)}{z_1-\ov{z}_3}+\frac{(\widetilde{\varepsilon}_{3}  \cdot k_2)}{z_2-\ov{z}_3}\Bigg)\nonumber\\
& +\frac{(\ov{\theta}_2 \theta_3 \varepsilon_2 \cdot q)  (\ov{\theta}_3 \ov{\theta}_1 \varepsilon_3 \cdot \varepsilon_1)}{(z_3-z_1)(z_2-z_3)}\theta_4 \ov{\theta}_4\Bigg( \frac{(\widetilde{\varepsilon}_{3}  \cdot k_1)}{z_1-\ov{z}_3}+\frac{(\widetilde{\varepsilon}_{3}  \cdot k_2)}{z_2-\ov{z}_3}\Bigg)
\label{openup}\\
&  +\frac{(\ov{\theta}_2 \ov{\theta}_1 \varepsilon_2 \cdot \varepsilon_1)}{(z_2-z_1)}\theta_3 \ov{\theta}_3\Bigg( \frac{(\varepsilon_{3}  \cdot k_1)}{z_1-z_3}+\frac{(\varepsilon_{3}  \cdot k_2)}{z_2-z_3}\Bigg) \theta_4 \ov{\theta}_4\Bigg( \frac{(\widetilde{\varepsilon}_{3}  \cdot k_1)}{z_1-\ov{z}_3}+\frac{(\widetilde{\varepsilon}_{3}  \cdot k_2)}{z_2-\ov{z}_3}\Bigg) 
\Bigg\}\ .\nonumber
 \end{align} 
Many of the possible terms vanish due to the on--shell conditions (\ref{shell}). In fact, using momentum conservation  and performing the Gra\ss mann integrals we arrive at
\be
 \mathcal{I}(2;1)=C\;    \frac{(\widetilde{\varepsilon}_3 \cdot k_1)}{|(z_1-z_3)|^2|(z_2-z_3)|^2}\Bigg\{(\varepsilon_2 \cdot \varepsilon_3)(\varepsilon_1 \cdot q)-(\varepsilon_1 \cdot \varepsilon_3)(\varepsilon_2 \cdot q)-(\varepsilon_1 \cdot \varepsilon_2)(\varepsilon_3 \cdot k_2)\Bigg\}\ , \label{2}
    \ee
    which agrees with \req{2a}.
The result \req{2} may be compared with the  four open superstring 
object following from \req{twistedWitten} for $n\!=\!4$. However, the main differences in the case at hand originate from applying  the on--shell condition \req{shell} which discards all the terms proportional to $p_i p_j$ and leads to \req{2}. Furthermore,  due  
to $\varepsilon^\mu_{\  \mu}=0$ in \req{2} there  is not any contribution from the bosonic contraction $\langle\partial X \partial X \rangle$.

We now rewrite \req{2} in terms of matrix notation and  Gra\ss mann integrations.
For this we introduce the $8\times 8$ matrix of block structure 
\be\label{matrix}
\Psi^{(2, 1)}=
\Psi_3\otimes\Psi_1\ ,
\ee
which splits into the $6\times 6$ block
 \begin{equation}\small{
\begin{aligned}
 & \hspace{-0.25cm}\Psi_3=\begin{pmatrix} 
 0 & 0 & 0 & \frac{-\varepsilon_1 \cdot q}{z_1-z_3}-\frac{\varepsilon_1 \cdot k_2}{z_1-z_2} & \frac{-\varepsilon_2 \cdot k_1}{z_2-z_1} & \frac{-\varepsilon_3 \cdot k_1}{z_3-z_1} 
 \\ 0 & 0 &  0 & \frac{-\varepsilon_1 \cdot k_2}{z_1-z_2} & \frac{-\varepsilon_2 \cdot k_1}{z_2-z_1}-\frac{\varepsilon_2 \cdot q}{z_2-z_3} & \frac{-\varepsilon_3 \cdot k_2}{z_2-z_3} 
 \\ 0 & 0 &  0 & \frac{-\varepsilon_1 \cdot q}{z_1-z_3} & \frac{-\varepsilon_2 \cdot q}{z_2-z_3} & \frac{-\varepsilon_3 \cdot k_1}{z_3-z_1}-\frac{\varepsilon_3 \cdot k_2}{z_3-z_2}
 \\ \frac{\varepsilon_1 \cdot q}{z_1-z_3}+\frac{\varepsilon_1 \cdot k_2}{z_1-z_2} & \frac{\varepsilon_1 \cdot k_2}{z_1-z_2} &  \frac{\varepsilon_1 \cdot q}{z_1-z_3} & 0 & \frac{\varepsilon_1 \cdot \varepsilon_2}{z_1-z_2} & \frac{\varepsilon_1 \cdot \varepsilon_3}{z_1-z_3}
 \\ \frac{\varepsilon_2 \cdot k_1}{z_2-z_1} & \frac{\varepsilon_2 \cdot k_1}{z_2-z_1}+\frac{\varepsilon_2 \cdot q}{z_2-z_3} &  \frac{\varepsilon_2 \cdot q}{z_2-z_3} & -\frac{\varepsilon_1 \cdot \varepsilon_2}{z_1-z_2} & 0 & \frac{\varepsilon_2 \cdot \varepsilon_3}{z_2-z_3}
 \\ \frac{\varepsilon_3 \cdot k_1}{z_3-z_1} & \frac{\varepsilon_3 \cdot k_2}{z_2-z_3} &  \frac{\varepsilon_3 \cdot k_1}{z_3-z_1}+\frac{\varepsilon_3 \cdot k_2}{z_3-z_2} & -\frac{\varepsilon_1 \cdot \varepsilon_3}{z_1-z_3} & -\frac{\varepsilon_2 \cdot \varepsilon_3}{z_2-z_3} & 0
  \end{pmatrix}
  \end{aligned} }\label{Psi3}
\end{equation}
and the following $2\times 2$ block:
\begin{equation}\label{Psi1}
 \Psi_1=\begin{pmatrix}
  0 & \frac{\widetilde{\varepsilon}_3 \cdot k_1\ov{z}_{21}}{\ov{z}_{13}\ov{z}_{23}} 
 \\  -\frac{\widetilde{\varepsilon}_3 \cdot k_1 \ov{z}_{21}}{\ov{z}_{13}\ov{z}_{23}} & 0 
  \end{pmatrix}\ .
  \end{equation}
We have defined the matrix \req{matrix} as concatenation in order to properly describe
the action of the Gra\ss mann matrix notation. More precisely, we define:
\be
\sum\limits_{i,j=1}^4 (\theta_j \,\, \bar{\theta}_j) \,\, \Psi^{(2;1)}\begin{pmatrix} \theta_i \\ \bar{\theta}_i  \end{pmatrix}:=
\sum\limits_{i,j=1}^3 (\theta_j \,\, \bar{\theta}_j) \Psi_3   \begin{pmatrix} \theta_i \\ \bar{\theta}_i  \end{pmatrix}+
 (\theta_4 \,\, \bar{\theta}_4) \Psi_1 \begin{pmatrix} \theta_4 \\ \bar{\theta}_4 \end{pmatrix} \ .\label{nested}
\ee
With these preparations and applying the well--known formula for the Gra\ss mann integrals 
\begin{equation}
\int \prod_{i=1}^m d \theta_i d\ov{\theta}_i\ \exp\Bigg\{\sum\limits_{i,j=1}^m (\theta_j \,\, \bar{\theta}_j) M\begin{pmatrix} \theta_i \\ \bar{\theta}_i  \end{pmatrix}\Bigg\}=\pf\; M\ ,
\end{equation}
with $M$ being a $2m\times 2m$ matrix, we can express \req{2} as:
   \begin{align}
   \mathcal{I}(2;1)&= \int  \prod\limits_{i=1}^{4}  \frac{\theta_1 \theta_2}{z_1-z_2} d \theta_i d\bar{\theta_i}  \exp\Bigg\{\ap^2\sum\limits_{i,j=1}^4 (\theta_j \,\, \bar{\theta}_j) \Psi^{(2;\mathbf{1})}  \begin{pmatrix} \theta_i \\ \bar{\theta}_i  \end{pmatrix}\Bigg\}\nonumber\\
  &= \int  \prod\limits_{i=1}^{4}  \frac{\theta_1 \theta_2}{z_1-z_2} d \theta_i d\bar{\theta_i}  \exp\Bigg\{\ap^2\; (\theta_3 \,\, \bar\theta_1\,\,\bar\theta_2\,\,\bar{\theta}_3) \Psi^{12}_{3} \begin{pmatrix} \theta_3 \\ \bar{\theta}_1 \\ \bar\theta_2 \\ \bar\theta_3 \end{pmatrix} +\ap^2\; \theta_4 \ov{\theta}_4\; \frac{\widetilde{\varepsilon}_3 \cdot k_1\ov{z}_{21}}{\ov{z}_{13}\ov{z}_{23}} \Bigg\}\nonumber\\
  & =\pf \Psi_{1}\; \fc{\pf\Psi^{12}_{3}}{z_1-z_2}=\pf \Psi_{1} \; \pf'\Psi_{3}\ . \label{const}
  \end{align}
According to the definition \req{redPF} the prime at the Pfaffian accounts for the additional factor $\tfrac{1}{z_1-z_2}$ in the integral and we have used the definition of $\pf \Psi_1$ as shown in \ref{A1}:
\begin{equation}
\pf \Psi_1= (\varepsilon_3  k_1)\; \frac{\ov{z}_{12}}{\ov{z}_{13}\ov{z}_{23}}\ .
\end{equation}
Above, we again  used that we may set $\widetilde{\varepsilon}_3=\varepsilon_3$. 
Note, that the anti--holomorphic part of the graviton vertex \req{Grav1} and hence half of its fermionic variables $(\theta_4,\ov \theta_4)$ are nested  into the   block diagonal matrix $\Psi_1$, which is concatenated with the rest $\Psi_3$, cf. also \req{nested}.
As a consequence  this procedure yields a different Gra\ss mannian structure than in the pure open superstring case \req{twistedWitten}.

Let us make some comments. Firstly, the expression \req{const} is identical to the integrand of the disk amplitude \req{AMPEYM21}. This is due to $KN_{D_2}=1$ and our comment after eq. \req{2a} that we are dealing with an all order exact expression in $\ap$. 
All these properties are special due to the three--particle kinematics and the on-shell conditions
\req{shell}. As a consequence of all order exactness in $\ap$ the three--point amplitude \req{AMPEYM21}   behaves identical in the $\ap\ra0$ and $\ap\ra\infty$ limits, i.e. the integrand \req{const} is uniform in $\ap$.
Secondly, we have already pointed out after eq. \req{2a} that we are not concerned with the integrations of vertex positions over the string world--sheet. We may analytically continue the positions of the open string fields  $z_i$ and treat them  as complex coordinates  $z_1,z_2\in\IC$ living on an auxiliary sphere world--sheet.
Likewise the conformal Killing group is promoted from $SL(2,\IR)$  to 
$SL(2,\IC)$. To this end, the integrand  \req{const} is considered as  function  in the three complex coordinates on the complex sphere:
\be\label{gocomplex}
z_i,\in \IC\ \ \ ,\ \ \ i=1,\ldots,3\ .
\ee
This provides a natural embedding from the disk to the sphere, which we shall formulate  in  the  section \ref{Section4}. 
In fact, due to the special three particle kinematics the structure of the contractions in \req{const} is the same regardless of the embedding onto the sphere. We shall make some more comments on this after \req{Gen}. Finally, in the next subsection we shall see, that the CHY formalism yields the same result \req{const}.

\subsubsection{Result from CHY calculation}

To compare our result \req{const} with the EYM amplitude \req{AmpEYM} in the CHY formalism we determine the integrand \req{chy}
for the case of two gluons and one graviton:
\begin{equation}\label{chyEx}
    \mathcal{I}_{2+1}(1,2;q)=\mathcal{C}(1,2)\ C_{qq}\ \pf'\Psi_{3}(k_a,q,\varepsilon,\sigma)\ . 
\end{equation}
Therefore, we need to compute the reduced Pfaffian $\pf'\Psi_{3}$ and the $C_{qq}$. With $q$ denoting the third leg the latter is given by \req{cqq}:
\begin{equation}
   C_{qq}\equiv\pf \Psi_{1}=\sum\limits_{l=1}^{1} (\varepsilon_3  x_l)\ \frac{\sigma_{l,l+1}}{\sigma_{l,3}\sigma_{l+1,3}}=(\varepsilon_3  k_1)\ \frac{\sigma_{1,2}}{\sigma_{1,3}\sigma_{2,3}}\ .
\end{equation}
The color form (2.35) becomes $\mathcal{C}(1,2)=(\sigma_1-\sigma_2)^{-2}$. On the other hand, the corresponding matrix $\Psi_{3}$ can be determined from \req{psi2} by taking into account that for this amplitude all kinematical invariants $p_ip_j$ are zero. It leads to the same matrix already defined in \req{Psi3} with all $z_i$ replaced by $\si_i$.
Removing from the latter matrix the first and second rows and columns gives $ \Psi^{12}_{3}$.
Eventually, the amplitude \req{chyEx} becomes:
\begin{equation}
\begin{aligned}
   \mathcal{I}_{2+1}(1,2;1) &=\mathcal{C}(1,2)\ (\varepsilon_3 \cdot k_1) \frac{\sigma_1-\sigma_2}{(\sigma_1-\sigma_3)\;(\sigma_2-\sigma_3)}\ \frac{\pf\Psi^{12}_{3}}{\sigma_1-\sigma_2}  \\
    &=\frac{1}{(\sigma_1-\sigma_2)^2}\ \frac{(\varepsilon_3 \cdot k_1)}{(\sigma_1-\sigma_3)^2(\sigma_2-\sigma_3)^2}\\
    &\times\Bigg\{(\varepsilon_2 \cdot \varepsilon_3)(\varepsilon_1 \cdot q)-(\varepsilon_1 \cdot \varepsilon_3)(\varepsilon_2 \cdot q)+(\varepsilon_1 \cdot \varepsilon_2)(\varepsilon_3 \cdot k_1)\Bigg\}\ .
    \end{aligned}\label{plitude}
\end{equation}
Furthermore, for the three--point massless amplitude the conditions from the scattering equations (\ref{f}) are trivially satisfied since all kinematical invariants $p_i  p_j$ are zero. To this end, for the case at hand we can write \req{AmpEYM} as:
\begin{align}
    \mathcal{A}_{CHY}(2;1)&=\int \frac{d \sigma_1d \sigma_2 d \sigma_3}{\textit{SL}(2,\IC)} \frac{(\varepsilon_3 \cdot k_1)}{(\sigma_1-\sigma_2)^2(\sigma_1-\sigma_3)^2(\sigma_2-\sigma_3)^2}\nonumber\\
    &\times\Bigg\{(\varepsilon_2 \cdot \varepsilon_3)(\varepsilon_1 \cdot q)-(\varepsilon_1 \cdot \varepsilon_3)(\varepsilon_2 \cdot q)+(\varepsilon_1 \cdot \varepsilon_2)(\varepsilon_3 \cdot k_1)\Bigg\}\ . \label{aamplitude}
\end{align}
The kinematical part of \req{aamplitude} agrees with (\ref{2a}) stemming  from the superstring calculation.

\subsection{Construction of the twisted form $\varphi^{EYM}_{2;{\bf 1}}$}
\label{construct21}

Equipped with the results from the previous subsections here we want to propose a pair of appropriate twisted forms $\varphi_+$ and $\varphi_-$  to express the EYM amplitude (\ref{aamplitude}) involving two gluons and one graviton   by an appropriate intersection number \req{intersection}. Then, the CHY amplitude (\ref{aamplitude}) is found in the leading $\ap\ra\infty$ limit of the latter as \req{expansion}:
\be\label{Setup21}
\Ac_{CHY}(2;1)=\lim_{\ap\ra\infty}\vev{\varphi_+ , \varphi_-}_\omega\ .
\ee
In the following we shall motivate the construction of our two twisted forms by inspecting the direct sum structure \req{nested}, which furnishes the following direct product structure  of two Gra\ss mann integrals
  \begin{equation}
    \begin{aligned}
 \mathcal{I}(2;1) & = \int  \prod\limits_{i=1}^{3}  \frac{\theta_1 \theta_2}{z_1-z_2} \ d \theta_i d\bar{\theta_i} \ \exp\Bigg\{\ap^2 \sum\limits_{i,j=1}^3 (\theta_j \,\, \bar{\theta}_j) \Psi_3   \begin{pmatrix} \theta_i \\ \bar{\theta}_i  \end{pmatrix} \Bigg\}  \\
  &\times \int d\theta_{4} d \ov{\theta}_4\ \exp\Bigg\{
   \ap^2\; (\theta_4 \,\, \bar{\theta}_4) \Psi_1 \begin{pmatrix} \theta_4 \\ \bar{\theta}_4 \end{pmatrix} \Bigg\}\ KN\cdot \overline{KN}\ , \label{KLTstruct}
  \end{aligned}
  \end{equation}
with the matrices $\Psi_3$ and $\Psi_1$ given in \req{Psi3} and \req{Psi1}, respectively.  Above we have appended  Koba--Nielsen factors, which of course are 
trivial $KN_{D2}\!=\!|KN|^2\!=~\!1$ due to  \req{KNtrivi}.
The expression \req{KLTstruct} reminds of a KLT like product, which connects a  holomorphic and anti--holomorphic  sector yet without any color ordering.
We will associate the first holomorphic factor of \req{KLTstruct} to be an element of $H^{r+n-3}_{+\omega}$. On the other hand, by using the isomorphy \req{ISO} between dual twisted cohomologies we may
associate the second anti--holomorphic  factor of \req{KLTstruct} with a twisted  form from $H^{r+n-3}_{-\omega}$
by imposing the map:
\begin{align}
 \ov{z_i} &\longrightarrow z_i\ \ \ ,\ \ i=1,\ldots,n+r\ ,\nonumber\\
 \overline{KN} &\longrightarrow KN^{-1}\ .\label{ISOMAP}
\end{align}
With these preparations as our first twisted form we can choose:
\be
 \varphi^{EYM}_{\pm,2;1}=d\mu_{3} \ \int  \prod\limits_{i=1}^{3}  \frac{\theta_1 \theta_2}{z_1-z_2}\ d \theta_i d\bar{\theta_i} \ \exp\Bigg\{\ap^2 \sum\limits_{i,j=1}^3 (\theta_j \,\, \bar{\theta}_j) \Psi_3   \begin{pmatrix} \theta_i \\ \bar{\theta}_i  \end{pmatrix} \Bigg\}\ .\label{pht}
\ee  
As the second twisted form we now choose
 \be\label{pht1}
\widetilde{\varphi}^{EYM}_{\pm,2;1}=d\mu_{3} \,\,\mathcal{C}(1,2)\int d \theta_4 d\bar{\theta_4}\   \exp\Bigg\{ \ap^2 (\theta_4 \,\, \bar{\theta}_4) \Psi_1\begin{pmatrix} \theta_4 \\ \bar{\theta}_4 \end{pmatrix} \Bigg\}\Bigg|_{\ov{z_l} \rightarrow z_l} \ ,
  \ee
subject to \req{ISOMAP}, which entails: 
\begin{equation}
\Psi_1\big|_{\ov{z}_l \rightarrow z_l}=\begin{pmatrix}
  0 & \frac{\varepsilon_3 \cdot k_1z_{21}}{z_{13}z_{23}} 
 \\  -\frac{\varepsilon_3 \cdot k_1 z_{21}}{z_{13}z_{23}} & 0 
  \end{pmatrix} \ , \qquad\qquad  \pf \Psi_1\big|_{\ov{z_l} \rightarrow z_l}= (\varepsilon_3  k_1)\; \frac{z_{12}}{z_{13}z_{23}}\ .
\end{equation}
In order to properly describe the color ordering of  \req{plitude} we have augmented \req{pht1} with  the following Parke--Taylor factor:
\be\label{CFO}
\mathcal{C}(1,2)=\frac{1}{(z_1-z_2)(z_2-z_1)}\ .
\ee
Eventually, computing the intersection number \req{intersection} of our two twisted forms \req{pht} and \req{pht1} yields
\begin{equation}
\begin{aligned}
    \mathcal{A}_{CHY}(2;1)&= \vev{\widetilde{\varphi}^{EYM}_{2;1} | \varphi^{EYM}_{2;1}}_\omega=\int \frac{d z_1 d z_2 d z_3}{\textit{SL}(2,\IC)}\ 
    \frac{(\varepsilon_3 \cdot k_1)}{(z_1-z_2)^2(z_1-z_3)^2(z_2-z_3)^2}\\[2mm]
    &\times \Bigg\{(\varepsilon_2 \cdot \varepsilon_3)(\varepsilon_1 \cdot q)-(\varepsilon_1 \cdot \varepsilon_3)(\varepsilon_2 \cdot q)+(\varepsilon_1 \cdot \varepsilon_2)(\varepsilon_3 \cdot k_1)\Bigg\} \ ,\label{amplitude}
    \end{aligned}
\end{equation} 
which is the CHY  amplitude given in (\ref{aamplitude}).
Actually, in the case at hand the $\ap\ra\infty$ limit \req{expansion} is exact in $\alpha'$, i.e.:
\be
\Ac_{CHY}(2;1)=\lim_{\ap\ra\infty}\; \vev{\widetilde{\varphi}^{EYM}_{2;1}| \varphi^{EYM}_{2;1}}_\omega\ .
\ee
Not any  lower orders in $\ap$ show up due to the limited number of contractions for this case.

\sect{Embedding of the disk onto the sphere}\label{Section4}

The ambitwistor string has been defined as a chiral string theory whose tree amplitudes reproduce the CHY formulae for massless scattering \cite{Mason}. 
By construction it only contains left--moving world--sheet fields and has no massive states. It can be described as the $\ap\!\ra\!0$ limit  of superstring theory.

In contrary, in this work we take the conventional superstring theory to provide twisted forms which are suitable to describe EYM amplitudes in their $\ap\!\ra\!\infty$ limit.
The latter is also the limit, where intersection theory provides the standard CHY formulae for
pure gauge and gravity amplitudes \cite{MT}. While the selection of the relevant twisted form
\req{Witten} to describe pure gauge and gravity amplitudes follows from an anlog expression in open superstring  theory  \req{twistedWitten} the description of the EYM amplitudes is different due to the presence of both gluons and gravitons. In fact, in string theory such an amplitude is described by a superstring disk amplitude involving both open and closed strings \cite{stie}. Due to the boundary of the disk world--sheet interactions between holomorphic and anti--holomorphic closed string fields appear. This mixing has to be taken into account when 
building the twisted forms relevant for EYM amplitudes.
In this section we prepare the necessary steps for constructing the latter.
In particular, we shall promote the disk correlator onto a larger space by some field extension of the open string vertex operator. This map promotes the string theory on the disk to a holomorphic theory on the sphere.

\subsect{Disk embedding}

In section \ref{Section3} we have derived our EYM amplitude \req{amplitude} by considering a disk amplitude
and extracting the pair of twisted form \req{pht} and \req{pht1}. On the other hand, in the setup for  ambitwistor strings and  intersection theory all results are derived on the sphere \cite{CHYprl,Mason,M2}.  In fact, in order to proceed to the multi--leg case within our  construction  we shall establish an embedding of the disk amplitude onto the sphere. In lines of  \req{gocomplex} this requires promoting  the fields of the open string vertex operators \req{Open0} and \req{Open1}, which are defined over the real line, to the full complex plane and analyze their holomorphic and anti--holomorphic properties. In order to accomplish  this we shall look at their equations of motions:
\begin{equation}
   \begin{aligned}
     \partial \ov{\partial}\widetilde{X}^\mu &=0\ , \\
      \ov{\partial}\widetilde{\psi}^\mu&=0\ ,\\
      \partial \widetilde{\psi}^\mu& =0\ .
    \end{aligned}
\end{equation}
We can define an embedding of the $n$ open string vertex positions $x_i\in\IR$ located on the boundary of the disk onto the the sphere $z_i\in\IC$ as:
\begin{equation}\label{Embedding}
 x_i\longmapsto z_i\ \ \ ,\ \ \ i=1,\ldots,n\ .
\end{equation}
Similarly, we define an embedding of  the  open string fields on the disk onto the sphere compatible with the equations of motion as: 
\begin{equation}
   \begin{aligned}
       X^\mu(x) & \longmapsto X^\mu(z)+\widetilde{X}^\mu(\ov{z}) \ \Rightarrow \ \begin{cases}  \partial X& \mapsto \partial X \\
                                                                                    & k \cdot X \mapsto k \cdot (X+\widetilde{X})\ , \end{cases}  \\
       \psi^\mu(x)  &\longmapsto\psi^{\mu}(z) \ .    \end{aligned} \label{emb}
\end{equation}
In addition, to adopt a given color ordering from the disk an anti--holomorphic gauge current $\tilde J^c(\bar z)$ is appended to each open string vertex operators.
In total, the  string vertex operators on the sphere will be defined as: 
\begin{equation}
   \begin{aligned}
  V_o(\varepsilon_i,k_i,x_i) &\longmapsto V_o(\varepsilon_i,k_i,z_i)\ e^{ik_i \widetilde{X}(\ov{z}_i)}\ \tilde J^{c_i}(\bar z_i)\ ,\ \ &i=1,\ldots,n\ ,\\[2mm]
   V_c(\varepsilon_s,q_s,z_{n+s},\ov{z}_{n+s}) &\longmapsto V_c=V_o(\varepsilon_s,q_s,z_{n+s})\ V_o(\widetilde{\varepsilon}_s,\widetilde{q}_s,\ov{z}_{n+s}) \ ,\ \ &s=1,\ldots, r\ .
    \end{aligned}\label{MAP}
\end{equation}
Thus, the open string vertex operators acquire an additional anti--holomorphic field dependence, while the closed string vertex operators remain untouched.
In this embedding we also promote the conformal Killing group from $SL(2,\IR)$ to $SL(2,\IC)$  and the full disk  correlator $\vev{\ldots}_{D_2}$ has to be the treated as if it was defined on the sphere $\vev{\ldots}_{S_2}$. I.e. instead \req{genz} we shall now consider the product of correlators on the double cover $S_2$:
\begin{align}
\Big\langle \prod\limits_{i=1}^{n} &V_o(\varepsilon_i,k_i,z_i)\ \prod\limits_{s=1}^{r} V_c (\varepsilon_s,q_s,z_{n+s},\bar{z}_{n+s})\Big\rangle_{D^2}  \label{Startn1}  \\
&\longmapsto \ \  \Big\langle \prod\limits_{i=1}^{n} V_o(\varepsilon_i,k_i,z_i)\ e^{i k_i \widetilde{X}(\ov{z}_i)}\ \prod\limits_{s=1}^{r} V_c (\varepsilon_s,q_s,z_{n+s},\bar{z}_{n+s})\Big\rangle_{S^2}\times 
\Big\langle \tilde J^{c_1}(\bar z_1)\ldots \tilde J^{c_n}(\bar z_n)\Big\rangle_{S^2}\ .\nonumber
\end{align}   
In particular, the map \req{MAP} changes the Koba--Nielsen factor  \req{KND2} to a correlator on the double cover~$S_2$
\be
KN_{D_2}\longmapsto\ KN_{S_2}\ ,
\ee
with
\be
KN_{S_2}=\prod_{i<j}^n |z_j-z_i|^{2\ap k_ik_j}\ \prod_{a<b}^{r}|z_{n+b}-z_{a+n}|^{2\ap q_aq_b}
\ \prod_{i=1}^n\prod_{a=1}^{r} |z_{a+n}-z_i|^{2\ap k_iq_a}\equiv KN\cdot \overline{KN}\ ,
\ee
which in turn can be split into a pair of  pure holomorphic and anti--holomorphic factors
 \begin{equation}
    \begin{aligned}
KN&= \prod_{i<j}^n (z_j-z_i)^{\ap k_ik_j}\ \prod_{a<b}^{r}(z_{n+b}-z_{a+n})^{\ap q_aq_b}
\ \prod_{i=1}^n\prod_{a=1}^{r} (z_{a+n}-z_i)^{\ap k_iq_a}\\
&= \prod_{i<j}^{n+r}(z_j-z_i)^{\ap p_ip_j}=:e^{\int_\gamma\omega}\ ,\\
\overline{KN}&=\prod_{i<j}^n (\ov z_j-\ov z_i)^{\ap k_ik_j}\ \prod_{a<b}^{r}(\ov z_{n+b}-\ov z_{a+n})^{\ap q_aq_b}\ \prod_{i=1}^n\prod_{a=1}^{r} (\ov z_{a+n}-\ov z_i)^{\ap k_iq_a}\\
&=\prod_{i<j}^{n+r}(\bar z_j-\bar z_i)^{\ap p_ip_j}=:e^{\int_{\bar\gamma}\bar\omega}\ ,   
\end{aligned}\label{KNs}
\end{equation}
respectively.
Similar to \req{KN} in  \req{KNs} we have defined the one--forms $\omega$ and $\bar\omega$ specifying the twisted cohomologies $H^{n+r-3}_{\omega}$ and $H^{n+r-3}_{\overline\omega}$, respectively.

It is worth pointing out the resemblance of our embedding to the construction of heterotic string theory in which one has a superstring sector for right--movers and a bosonic string sector for the left--movers. However, one crucial difference is that we are 
dealing with a left and right supersymmetric closed string sector and only 
 extend the real variables of the  open string. 
It would be interesting to find connections between our construction and that of the  
heterotic ambitwistor string.

\subsect{Sphere integrand from the disk embedding}

Let us summarize our procedure how to extract twisted forms  for EYM amplitudes from string disk amplitudes $\mathcal{A}(n;r)$. The latter  involve $n$ open and $r$ closed strings described by  the vertex operators given in subsection \ref{vertexO}.
Actually,  since we are interested in the construction of twisted forms we suppress the complex integral over the vertex operator positions and consider only the disk correlator
\req{genz}, which  computes the integrand $\mathcal{I}(n;r)$ for the EYM string amplitude $\mathcal{A}(n;r)$. 
Now, we reformulate  the disk correlator \req{genz} as a sphere correlator \req{Startn1}
by extending the open string fields onto the \req{MAP}. The specific color ordering of the open strings
along the disk boundary is adopted by the gauge current correlator. The gauge correlator $\langle \tilde J^{c_1}(\bar z_1)\ldots \tilde J^{c_n}(\bar z_n)\rangle_{S^2}$ decomposes into a sum over various gauge group structures. Since we are only concerned with  a single trace color structure of the form $\Tr(T^{c_1}\ldots T^{c_n})$ we shall project the gauge current correlator onto the relevant color form \req{ColorForm}
\be
\Big\langle \tilde J^{c_1}(\bar z_1)\ldots \tilde J^{c_n}(\bar z_n)\Big\rangle_{S^2}\ \longrightarrow\   \Tr(T^{c_1}\ldots T^{c_n})\ \mathcal{C}(1,2,\ldots,n)\ ,
\ee
with:
\be
 \mathcal{C}(1,2,\ldots,n) = \frac{1}{(\bar z_1- \bar z_2)(\bar z_2-\bar z_3)\cdot\ldots\cdot(\bar z_n-\bar z_1)}\ .\label{TWISTN1}
\end{equation}

As next step we  analyze  the field  contractions of the correlator \req{Startn1} and 
express the field interactions in terms of the Gra\ss mann formalism \req{twistedWitten}.
This leads to the following integrand: 
\begin{equation}
    \begin{aligned}
\mathcal{I}(n;r) &  = \mathcal{C}(1,2,\ldots,n)\ \times\Big\langle \,\,\, \int  (\prod\limits_{i=1}^{n+2r}d\theta_i d\bar{\theta}_i) \ \frac{\theta_1 \theta_2}{z_1-z_2} \\
      &\times \exp\lf\{ik_1\cdot (X+\widetilde{X})+\theta_1\bar{\theta_1} \varepsilon_1 \cdot \partial X+\theta_1 \sqrt{\alpha'} k_1\cdot \psi+\bar{\theta_1} \sqrt{\alpha'} \varepsilon_1 \cdot \psi\ri\} \\
     &\hskip6cm \vdots \\
     &\times \exp\lf\{ik_n\cdot (X+\widetilde{X})+\theta_{n}\bar{\theta}_{n} \varepsilon_n \cdot \partial X+\theta_{n} \sqrt{\alpha'} k_n\cdot \psi+\bar{\theta}_{n} \sqrt{\alpha'} \varepsilon_n \cdot \psi\ri\} \\
     &\times \exp\lf\{i q_1 \cdot X+i \widetilde{q}_1 \cdot \widetilde{X}+\theta_{n+1}\bar{\theta}_{n+1} \varepsilon_{n+1} \cdot \partial X+\theta_{n+1} \sqrt{\alpha'} q_1\cdot \psi+\bar{\theta}_{n+1} \sqrt{\alpha'} \varepsilon_{n+1} \cdot \psi\ri.\\
     &\hskip2cm \lf.+\theta_{n+2}\bar{\theta}_{n+2} \widetilde{\varepsilon}_{n+1} \cdot \ov{\partial} \widetilde{X} +\theta_{n+2} \sqrt{\alpha'} \widetilde{q}_1\cdot \widetilde{\psi}+\bar{\theta}_{n+2} \sqrt{\alpha'} \widetilde{\varepsilon}_{n+1} \cdot \widetilde{\psi}
      \ri\} \\
  &\hskip6cm \vdots \\
   & \times\exp\lf\{i q_{r} \cdot X+i \widetilde{q}_{r} \cdot \widetilde{X}+\theta_{2r+n-1}\bar{\theta}_{2r+n-1} \varepsilon_{n+r} \cdot \partial X+\theta_{2r+n-1}\sqrt{\alpha'}  q_{r}\cdot \psi+\bar{\theta}_{2r+n-1} \sqrt{\alpha'} \varepsilon_{n+r} \cdot \psi\ri.\\
   & \hspace{2cm}+\lf.\theta_{2r+n}\bar{\theta}_{2r+n} \widetilde{\varepsilon}_{n+r} \cdot \ov{\partial}\widetilde{X} +\theta_{2r+n} \sqrt{\alpha'}  \widetilde{q}_{r}\cdot \widetilde{\psi}+\bar{\theta}_{2r+n} \sqrt{\alpha'} \widetilde{\varepsilon}_{n+r} \cdot \widetilde{\psi}   \ri\}\Big\rangle_{S_2}\ .
     \end{aligned}\label{IN1}
\end{equation}
Again, the  background ghost charge is taken into account by the term $\tfrac{\theta_1 \theta_2}{z_1-z_2}$. 
The integrand \req{IN1} can be further simplified such that the fermionic variables $\theta_i\bar\theta_i$ resemble the pattern dictated by the map \req{MAP}.
Furthermore, the KLT like product structure \req{KN} connecting the holomorphic and
anti--holomorphic sector will be exposed, which then will allow us to read off the relevant twisted forms by using the isomorphism \req{ISOMAP}.
As a last  step, the  $\alpha'\ra\infty$ can be taken. Following (\ref{expansion}) this yields  the integral localization over the scattering equations.

\sect{Twisted form  and intersections for EYM amplitudes with one graviton}\label{Section5}

In this section we perform the first steps towards extending our result \req{amplitude} for two gluons and one graviton to the generic case by  increasing the number of gluons, i.e. open strings in the underlying string correlator \req{Startn1}. 
To simplify the Gra\ss mann integral \req{IN1} we perform similar steps, which took us from \req{step1} to \req{const}. In addition,  following subsection \ref{Notation} for the open and closed string states we introduce the unifying notation (subject to $\varepsilon_{n+1}=\widetilde{\varepsilon}_{n+1}$): 
$$\begin{aligned}
  & \{\xi_1,...,\xi_{n+2}\}:=\{\varepsilon_1,\varepsilon_2,...,\varepsilon_n,\varepsilon_{n+1},\widetilde{\varepsilon}_{n+1}\}\ ,\\
  & \{p_1,...,p_{n+2}\}:=\{k_1,k_2...,k_n,q_1,\widetilde{q}_1\}\ .
\end{aligned}$$
With these preparations we can write \req{IN1} as
    \begin{equation}
    \begin{aligned}
     \mathcal{I}(n;1)&=\mathcal{C}(1,2,...,n)\  \int \prod\limits_{i=1}^{n+2} d \theta_i d\bar{\theta_i}\ \frac{\theta_1 \theta_2}{z_1-z_2}\\
     & \times \exp\Bigg\{\ap^2\Bigg(
     \sum\limits_{i,j=1\atop i\neq j}^{n+1} \frac{ (\bar{\theta}_i \theta_j \xi_i \cdot p_j)}{z_i-z_j \mp \alpha'^{-1} \theta_i \theta_j}
  +\sum\limits_{i>j}^{n+1} \frac{( \bar{\theta_i} \bar{\theta_j}\xi_i \cdot \xi_j)}{z_i-z_j \mp \alpha'^{-1} \theta_i \theta_j}+\sum\limits_{i,j=1\atop i\neq j}^{n+1}\frac{ (\theta_i \bar{\theta_i} \xi_i \cdot p_j)}{z_i-z_j \mp \alpha'^{-1} \theta_i \theta_j}\\
   &  +\sum\limits_{i>j}^{n+1} \frac{( {\theta_i} {\theta_j}p_i \cdot p_j)}{z_i-z_j \mp \alpha'^{-1} \theta_i \theta_j}+\sum\limits_{j=1}^{n}\frac{ (\theta_{n+2} \ov{\theta}_{n+2} \xi_{n+2} \cdot p_j)}{\ov{z}_{n+1}-\ov{z}_j \mp \alpha'^{-1} \theta_{n+2} \theta_j} \Bigg)\ \Bigg\}\ \times  KN\cdot \overline{KN}\ .  \label{Gen}
\end{aligned}
\end{equation}
Let us make some comments. Firstly, $KN\cdot \overline{KN}$ is the Koba--Nielsen factor  \req{KN} for $n+1$ closed strings. Secondly, comparing \req{Gen} with \req{const} we evidence that the main difference is the additional last term in the exponential of \req{Gen}. It accounts for the new type of interactions between the anti--holomorphic
 field $\bar\partial X$ from the single graviton vertex and the anti--holomorphic open string fields $e^{i k_i\tilde X(\bar z_i)}$ from the gluon vertices subject to the map \req{MAP}.
On the other hand, the disk amplitude \req{1}  involves additional interactions between holomorphic and anti--holomorphic fields which are accounted for by the additional terms produced by the map \req{MAP}. 
At any rate, due to special kinematics of  the three--point amplitude \req{shell}
to arrive at \req{const} we had not to apply the embedding \req{MAP}, but simply rewrote the disk integrand \req{step1} in terms of \req{twistedWitten}.
Nevertheless, applying the map \req{MAP} at \req{AMPEYM21} yields\footnote{In order to see this let us  look at the equation (\ref{Gen}), which for the case $n=2$ and $KN=1$ becomes: 
 \begin{equation}
    \begin{aligned}
     \mathcal{I}(2;1)&= \mathcal{C}(1,2)\ \int \prod\limits_{i=1}^{4} d \theta_i d\bar{\theta_i}\ \frac{\theta_1 \theta_2}{z_1-z_2}\ \exp\Bigg\{\ap^2\Bigg(
     \sum\limits_{i,j=1\atop i\neq j}^{3} \frac{ (\bar{\theta}_i \theta_j \xi_i \cdot p_j)}{z_i-z_j \mp \alpha'^{-1} \theta_i \theta_j}   +\sum\limits_{i>j}^{3} \frac{( \bar{\theta_i} \bar{\theta_j}\xi_i \cdot \xi_j)}{z_i-z_j \mp \alpha'^{-1} \theta_i \theta_j}\\
     &+\sum\limits_{i,j=1\atop i\neq j}^{3}\frac{ (\theta_i \bar{\theta_i} \xi_i \cdot p_j)}{z_i-z_j \mp \alpha'^{-1} \theta_i \theta_j}   +\sum\limits_{i>j}^{3} \frac{( {\theta_i} {\theta_j}p_i \cdot p_j)}{z_i-z_j \mp \alpha'^{-1} \theta_i \theta_j}+\sum\limits_{j=1}^{2}\frac{ (\theta_{4} \ov{\theta}_{4} \xi_{4} \cdot p_j)}{\ov{z}_{3}-\ov{z}_j \mp \alpha'^{-1} \theta_{4} \theta_j} \Bigg)\ \Bigg\}\ .
 \end{aligned}\label{kleinB}
\end{equation}
Expanding the exponential in \req{kleinB} w.r.t. the fermionic variables and using  the tracelessness condition of the graviton polarization gives:
 \begin{equation}
    \begin{aligned}
     \mathcal{I}(2;1) &  = \mathcal{C}(1,2)\  \int \prod\limits_{i=1}^{4}  \frac{\theta_1 \theta_2}{z_1-z_2} d \theta_i d\ov{\theta}_i \ \Bigg\{\frac{(\bar{\theta}_1 \theta_3 \varepsilon_1 \cdot q)  (\ov{\theta}_3 \bar{\theta_2} \varepsilon_3 \cdot \varepsilon_2) \ov{z}_{21}(\ov{\theta}_4 \theta_4 \widetilde{\varepsilon}_{3}  \cdot k_1)}{(z_1-z_3)(z_2-z_3)(\ov{z}_1-\ov{z}_3)(\ov{z}_2-\ov{z}_3)}\\
& +\frac{(\bar{\theta}_2 \theta_3 \varepsilon_2 \cdot q)  (\ov{\theta}_3 \ov{\theta}_1 \varepsilon_3 \cdot \varepsilon_1) \ov{z}_{21}(\bar{\theta_4} \theta_4 \widetilde{\varepsilon}_{3} \cdot k_1)}{(z_1-z_3)(z_2-z_3)(\ov{z}_1-\ov{z}_3)(\ov{z}_2-\ov{z}_3)}+\frac{(\ov{\theta}_1 \ov{\theta}_2 \varepsilon_1 \cdot \varepsilon_2) (\ov{\theta}_3 \theta_{3} \varepsilon_3 \cdot k_1) \ov{z}_{21}(\ov{\theta}_4 \theta_{4} \widetilde{\varepsilon}_{3}  \cdot k_1)}{(z_1-z_3)(z_2-z_3)(\ov{z}_1-\ov{z}_3)(\ov{z}_2-\ov{z}_3)} \Bigg\}\ .
\end{aligned}\label{kleinBB}
\end{equation}
Up to the color factor \req{CFO} the expression \req{kleinBB} agrees with \req{openup} subject to  the reality condition $z_1=\bar z_1$ and $z_2=\bar z_2$ of the two open string positions  on the disk.} the same result \req{const}.

Now we proceed similar as in subsection \ref{construct21} by making profit of the KLT like construction of EYM amplitudes established  by the map \req{MAP}.
We shall  construct from our integrand  \req{Gen} a pair of twisted forms $\varphi_+,\varphi_-$, whose intersection number  $\vev{\varphi_+|\varphi_-}_\omega$ will give in its $\ap\ra\infty$ limit \req{expansion} the EYM amplitude formula of the  CHY formalism  \req{AmpEYM}. 

For the case of $n$ gluons and one graviton  the latter yields the integrand (\ref{chy}):
\begin{equation}\label{chyeymN1}
    \mathcal{I}_{n+1}(1,2,\ldots,n;q)=\mathcal{C}(1,2,\ldots,n)\ C_{qq}\ \pf'\Psi_{n+1}(k_a,q,\varepsilon,\sigma) \ .
\end{equation}
Here $\mathcal{C}(1,2,\ldots,n)$ is the Parke--Taylor factor \req{ColorForm}.
The latter will be identified with the gauge correlator \req{TWISTN1} of the integrand \req{Gen}, which will be associated to the twisted form $\tilde\varphi^{EYM}_{\pm,n;1}$ in the same way as proposed in subsection \ref{construct21}.
To find the correct description of $C_{qq}\equiv \pf\Psi_{1}$  and $\pf'\Psi_{n+1}(k_a,q,\varepsilon,\sigma)$ we proceed as follows.
First note, that the denominators in the exponential proliferate terms subleading in $\ap$, 
in particular:
$$\frac{ \bar{\theta_i} \bar{\theta_j}(\xi_i \cdot \xi_j)}{z_i-z_j \mp \alpha'^{-1} \theta_i \theta_j}=\frac{ \bar{\theta_i} \bar{\theta_j}(\xi_i \cdot \xi_j)}{z_i-z_j }  \pm\alpha'^{-1}\ \frac{ \theta_i \theta_j \bar{\theta_i} \bar{\theta_j}(\xi_i \cdot \xi_j)}{(z_i-z_j)^2}\ .$$
With this information the integrand \req{Gen} can be cast into the following form:
\begin{equation}
\begin{aligned}
      \mathcal{I}(n;1)&=\mathcal{C}(1,2,\ldots,n)\  
      \int \prod\limits_{i=1}^{n+2} d \theta_i d\bar{\theta_i}\ \frac{\theta_1 \theta_2}{z_1-z_2}\ \ 
      \exp\lf\{\ap^2\;\psi_1\theta_{n+2} \ov{\theta}_{n+2}\ri\}\\
      &\times \exp\Bigg\{\h\alpha'^2\sum\limits_{i,j=1}^{n+1}\lf(\theta_i\atop\bar\theta_i\ri)^t
     \Psi_{n+1}
      \lf(\theta_j\atop\bar\theta_j\ri)\pm\h\alpha'\ \sum\limits_{i,j=1\atop i\neq j}^{n+1}\frac{ \theta_i \theta_j \bar{\theta_i} \bar{\theta_j}(\xi_i \cdot \xi_j)}{(z_i-z_j)^2}\Bigg\}\ \times  KN\cdot \overline{KN}\ .
     \end{aligned} \label{stc}
\end{equation}
In \req{stc} we have introduced   $(2n+2)\times (2n+2)$ matrix \req{PSI} 
\be
\Psi_{n+1}\equiv \Psi_{S_{n+1}}=\lf. \begin{pmatrix}
      A_{ij}&-C_{ji}\\
      C_{ij}&B_{ij}
      \end{pmatrix}\ri|_{\sigma_l=z_l}\ \ ,\ \ S_{n+1}=\{1,\ldots,n,n+1\}\ ,
\ee
with the $(n+1)\times (n+1)$ matrices $A,B$ and $C$ being the same as \req{psi2} with $\sigma_l$ replaced by $z_l,\; l\!=\!1,\ldots,n+1$. 
Furthermore, we have defined the $2\times 2$ matrix $\Psi_1$:
\be\label{PSI1}
\Psi_1=\begin{pmatrix}
      0&\psi_1\\
    -\psi_1&0
      \end{pmatrix}\ \ \ ,\ \ \ \psi_1=\sum_{j=1}^n\frac{(\xi_{n+2} p_j)}{\ov{z}_{n+1}-\ov{z}_j }\ .
\ee
With the help of \req{Psi} the expression \req{PSI1} can be cast into \req{cqq}.  Similar as in \req{matrix}  we can construct the $(2n+4)\times (2n+4)$ matrix $\Psi^{n,1}$ of holomorphic, anti holomorphic block structure
\be\label{Matrix}
\Psi^{(n;1)}=\Psi_{n+1}\otimes\Psi_1
\ee
to write the full $\ap^2$ order of the exponential of \req{stc} in terms of the concatenation: 
\be
\sum\limits_{i,j=1}^{n+2} (\theta_j \,\, \bar{\theta}_j) \,\, \Psi^{(n;1)}\begin{pmatrix} \theta_i \\ \bar{\theta}_i  \end{pmatrix}:=
\sum\limits_{i,j=1}^{n+1} (\theta_j \,\, \bar{\theta}_j) \Psi_{n+1}   \begin{pmatrix} \theta_i \\ \bar{\theta}_i  \end{pmatrix}+
 (\theta_{n+2} \,\, \bar{\theta}_{n+2}) \Psi_1 \begin{pmatrix} \theta_{n+2} \\ \bar{\theta}_{n+2} \end{pmatrix} \ .\label{Nested}
\ee
Eventually, with this block structure we are now able to construct our pair of twisted forms. We take the holomorphic part of \req{stc} (described by $\Psi_{n+1}$ and the subleading term) supplemented by the  Parke--Taylor factor \req{TWISTN1} to amount to the form 
$\varphi^{EYM}_{\pm,n;1}$:
\begin{equation}
    \begin{aligned}
   \varphi^{EYM}_{\pm,n;1}&=d\mu_{n+1} \,\,\int  \prod\limits_{i=1}^{n+1}  \frac{\theta_1 \theta_2}{z_1-z_2} d \theta_i d\bar{\theta_i}  \exp\Bigg\{\frac{1}{2}\ap^2 \sum\limits_{i,j=1}^{n+1} (\theta_j \,\, \bar{\theta}_j) \Psi_{n+1}  \begin{pmatrix} \theta_i \\ \bar{\theta}_i  \end{pmatrix} \Bigg\}\\
  &\times \exp \Bigg\{\pm\h\alpha'\ \sum\limits_{i,j=1\atop i\neq j}^{n+1}\frac{ \theta_i \theta_j \bar{\theta_i} \bar{\theta_j}(\xi_i \cdot \xi_j)}{(z_i-z_j)^2}\Bigg\}\ . 
  \end{aligned}\label{stcnew}
\end{equation}
Furthermore, we extract the anti--holomorphic part of \req{stc}, apply the isomorphism  \req{ISOMAP} to arrive at our  second twisted form  twisted form $\widetilde{\varphi}^{EYM}_{\pm,n;1}$:
\begin{equation}
  \widetilde{\varphi}^{EYM}_{\pm,n;\mathbf{1}} =d\mu_{n+1}\ \mathcal{C}(1,2,\ldots,n) \ \int d \theta_{n+2} d\bar{\theta}_{n+2}\  \exp\Bigg\{ \ap^2\; (\theta_{n+2} \,\, \bar{\theta}_{n+2}) \Psi_1\begin{pmatrix} \theta_{n+2} \\ \bar{\theta}_{n+2} \end{pmatrix} \Bigg\}\Bigg|_{\ov{z_l} \rightarrow z_l}.
  \label{stcnew1}
\end{equation}
The two twisted forms  \req{stcnew} and \req{stcnew1} are our candidate forms 
$\varphi_-=\widetilde{\varphi}^{EYM}_{\pm,n;1}$, $\varphi_+=\varphi^{EYM}_{-,n;1}$ for reproducing the EYM integrand \req{chyeymN1}. 
In fact, with \req{stcnew} and \req{stcnew1} we are now able to derive the EYM amplitude from \req{expansion}. The twisted forms in eq. \req{stcnew} and \req{stcnew1} are  of the appropriate form for taking the  $\alpha' \rightarrow \infty$ limit:
\begin{equation}
\begin{aligned}
\lim_{\ap\ra\infty}\hat{\varphi}^{EYM}_{\pm,n;1}&= \int  \prod\limits_{i=1}^{n+1}   
d \theta_i d\bar{\theta_i}\   \frac{\theta_1 \theta_2}{z_1-z_2}\ \exp\Bigg\{\frac{1}{2}\ap^2\sum\limits_{i,j=1}^{n+1} (\theta_j \,\, \bar{\theta}_j) \Psi_{n+1} \,\, \begin{pmatrix} \theta_i \\ \bar{\theta}_i  \end{pmatrix}\Bigg\}+\mathcal{O}(\ap^{-1})\\
     & =\fc{\pf\Psi^{12}_{n+1}}{z_1-z_2}+\mathcal{O}(\ap^{-1})= \pf' \Psi_{n+1}+\mathcal{O}(\ap^{-1})\ ,\\
   \lim_{\ap\ra\infty}\hat{\widetilde{\varphi}}^{EYM}_{\pm,n;1}&= \mathcal{C}(1,2,\ldots,n)\ \int  
d \theta_{n+2} d\bar{\theta}_{n+2}\  \ \exp\Bigg\{\ap^2\; (\theta_{n+2} \,\, \bar{\theta}_{n+2}) \Psi_1  \begin{pmatrix} \theta_{n+2} \\ \bar{\theta}_{n+2}  \end{pmatrix}\Bigg\}\Bigg|_{\ov{z_l} \rightarrow z_l}\\
& =  \mathcal{C}(1,2\ldots,n)\ \pf \Psi_{1}\big|_{\ov{z_l} \rightarrow z_l} \ .
\end{aligned}\label{TWISTEYMN1}
\end{equation}
Plugging the expressions \req{TWISTN1} and
\req{TWISTEYMN1} into \req{expansion}  yields the final expression for the EYM amplitude:
\begin{equation}
\begin{aligned}
 \mathcal{A}(1,2,\ldots,n;1)&=\lim_{\ap\ra\infty}
 \vev{\widetilde{\varphi}^{EYM}_{+,n;1},\varphi^{EYM}_{-,n;1}}_\omega
 =\int\limits_{\mathcal{M}_{0,n+1}}\!\!\! d \mu_{n+1}\ 
 \sideset{}{'}\prod_{a=1}^{n+1} \delta(f_a) \lim_{\alpha' \rightarrow \infty} 
  \hat{\varphi}_{n;1}^{EYM}\hat{ \widetilde{\varphi}}^{EYM}_{n;1}\\
 &=\!\!\!\!\int\limits_{\mathcal{M}_{0,n+1}}\!\!\! d \mu_{n+1}
 \sideset{}{'}\prod_{a=1}^{n+1} \delta(f_a)\;  \frac{\pf \Psi_{1}\big|_{\ov{z_l} \rightarrow z_l}\ \pf' \Psi_{n+1}}{(z_1-z_2)(z_2-z_3)\ldots(z_n-z_1)}\ .
 \end{aligned}\label{EXFORMM}
\end{equation}
In the leading order in $\ap$ this result agrees  with the expression given in \req{AmpEYM} for the Einstein Yang--Mills amplitude in the  CHY formalism. The terms $\Oc(\ap^{-1})$ subleading in $\ap$ originate from the limit \req{TWISTEYMN1}.

\sect{Twisted form  and intersections for general EYM amplitudes}

In this section we shall determine the pair of twisted forms $\varphi^{EYM}_{\pm,n;r}$ and $\widetilde{\varphi}^{EYM}_{\pm,n;r}$ for the  multi--graviton case.  To illuminate the structure and changes  compared to the one--graviton case we first  derive  the amplitude involving $n$ gluons and two gravitons. Equipped with these preparations  we then move to the generic case of $n$ gluons and $r$ gravitons.

\subsection{Twisted form and intersections for amplitudes of $n$ gluons and two gravitons} 

In this subsection we consider  the disk amplitude \req{genz} involving $n$ gluons and two gravitons. Subject to \req{MAP} we embed the latter onto the sphere and  arrive at \req{Startn1}. 
As already anticipated above  for the construction of the twisted differentials  we  do not need the position integrals and may focus on the integrand (\ref{IN1}) written in terms of Gra\ss mann variables. For the case at hand we introduce the unifying  notation for polarizations $\xi_i$, momenta $p_j$ and positions $\zeta_l$ exhibited in the  \ref{Notation}, more precisely (subject to $\tilde \eps_{n+i}=\eps_{n+i},\; i=1,\ldots,r$)
\begin{align}
  & \{\zeta_1,...,\zeta_{n+4}\}=\{z_1,z_2,...,z_n,z_{n+1},z_{n+1},z_{n+2},z_{n+2}\}\ ,\nonumber\\
  & \{\xi_1,...,\xi_{n+4}\}:=\{\varepsilon_1,\varepsilon_2,...,\varepsilon_n,\varepsilon_{n+1},\widetilde{\varepsilon}_{n+1},\varepsilon_{n+2},\widetilde{\varepsilon}_{n+2}\}\ ,\label{defsetss}\\
  & \{p_1,...,p_{n+4}\}:=\{k_1,k_2...,k_n,q_1,\widetilde{q}_1,q_2,\widetilde{q}_2\}\ ,\nonumber
\end{align}
and perform the field contractions using the correlators on the sphere. With these preparations the integrand (\ref{IN1}) assumes the following form:
  \begin{equation}
    \begin{aligned}
    \mathcal{I}(n;2) & =  \mathcal{C}(1,2,\ldots,n)\ \int \prod\limits_{i=1}^{n+4} d \theta_i d\bar{\theta_i} \ \frac{\theta_1 \theta_2}{\zeta_1-\zeta_2} \\
   &   \times\exp\Bigg\{\alpha'^2 \Bigg(\sum\limits_{{i,j=1\atop i\neq j} \atop i,j\neq n+2}^{n+3} \frac{ (\bar{\theta}_i \theta_j \xi_i \cdot p_j)}{\zeta_i-\zeta_j \mp \alpha'^{-1} \theta_i \theta_j}
     +\sum\limits_{{i,j=1\atop i>j} \atop i,j\neq n+2}^{n+3} \frac{( \bar{\theta_i} \bar{\theta_j}\xi_i \cdot \xi_j)}{\zeta_i-\zeta_j \mp \alpha'^{-1} \theta_i \theta_j}\\
     &\hskip0.5cm   +\sum\limits_{{i,j=1\atop i\neq j }  \atop i,j\neq n+2 }^{n+3}\frac{ (\theta_i \bar{\theta_i} \xi_i \cdot p_j)}{\zeta_i-\zeta_j \mp \alpha'^{-1} \theta_i \theta_j}  +\sum\limits_{{i,j=1\atop i>j} \atop i,j\neq n+2}^{n+3} \frac{( {\theta_i} {\theta_j}p_i \cdot p_j)}{\zeta_i-\zeta_j \mp \alpha'^{-1} \theta_i \theta_j}\\
   &\hskip0.5cm   +\sum\limits_{i,j\in \{n+2,n+4\} \atop i\neq j }\frac{ (\bar{\theta}_i \theta_j \xi_i \cdot p_j)}{\ov{\zeta}_i-\ov{\zeta}_j \mp \alpha'^{-1} \theta_i \theta_j}+\sum\limits_{ {j=1\atop  j\notin\{n+1,n+3\}} \atop i\in\{n+2,n+4\} }^{n+4}\frac{ (\theta_i \bar{\theta_i} \xi_i \cdot p_j)}{\ov{\zeta}_i-\ov{\zeta}_j \mp \alpha'^{-1} \theta_i \theta_j}\\
     &\hskip0.5cm     +\sum\limits_{j=n+2 \atop i=n+4} \frac{( \bar{\theta_i} \bar{\theta_j}\xi_i \cdot \xi_j)}{\ov{\zeta}_i-\ov{\zeta}_j \mp \alpha'^{-1} \theta_i \theta_j}+\sum\limits_{j=n+2 \atop i=n+4}\frac{( \theta_i \theta_j p_i \cdot p_j)}{\ov{\zeta}_i-\ov{\zeta}_j \mp \alpha'^{-1} \theta_i \theta_j} \Bigg)\Bigg\}\ \times   KN \cdot \overline{KN}\ .\label{Gen2}
\end{aligned}
\end{equation}
Note, that in \req{Gen2} the first four terms in the exponential account for the holomorphic
field contractions, while the last four terms represent the anti--holomorphic field contractions subject to the map \req{MAP}.
Furthermore, $KN\cdot \overline{KN}$ is the Koba--Nielsen factor  \req{KN} for $n+2$ closed strings.
In the following we explicitly outline the  steps leading to (\ref{Gen2}) and discuss the underlying matrix structure w.r.t. the fermionic variables. 
First of all, we discuss the anti--holomorphic sector of the two--graviton case. After factorizing  an $\alpha'^2$ factor we encounter two contributions. Firstly, there is the correlator of the  bosonic part  of one graviton vertex \req{Grav1} with both the open string exponentials stemming from the map \req{MAP} and a single  exponential from the second graviton (with $z_{l,m}=z_l-z_m$):
\begin{equation}
    \begin{aligned}
    \langle \widetilde{\varepsilon}^{n+1}_{\mu} \ov{\partial}  \widetilde{X}^{\mu} &(\ov{z}_{n+1})  \Bigg(\prod\limits_{i=1}^n e^{ik_i \widetilde{X}(\ov{z}_i)}\Bigg)\times e^{iq_{2}\widetilde{X}(\ov{z}_{n+2})} \rangle=i\Bigg(\sum\limits_{i=1}^{n} \frac{\widetilde{\varepsilon}^{n+1} \cdot k_i}{\ov{z}_i-\ov{z}_{n+1}}\Bigg)+i\frac{\widetilde{\varepsilon}^{n+1} \cdot q_2}{\ov{z}_{n+2}-\ov{z}_{n+1}}\\
     & =i\Bigg(\sum\limits_{i=1}^{n-1} \frac{\widetilde{\varepsilon}^{n+1} \cdot k_i}{\ov{z}_i-\ov{z}_{n+1}}\Bigg)+i \frac{\widetilde{\varepsilon}^{n+1} \cdot k_n}{\ov{z}_n-\ov{z}_{n+1}}+i\frac{\widetilde{\varepsilon}^{n+1} \cdot q_2}{\ov{z}_{n+2}-\ov{z}_{n+1}}\\
     & =i\sum\limits_{i=1}^{n-1} \frac{\widetilde{\varepsilon}^{n+1} \cdot k_i}{\ov{z}_i-\ov{z}_{n+1}}-i \frac{\widetilde{\varepsilon}^{n+1} \cdot \sum_{j=1}^{n-1}k_j+q_2}{\ov{z}_n-\ov{z}_{n+1}}+i\frac{\widetilde{\varepsilon}^{n+1} \cdot q_2}{\ov{z}_{n+2}-\ov{z}_{n+1}}\\
     & =i\sum\limits_{i=1}^{n-1} \frac{\widetilde{\varepsilon}^{n+1} \cdot k_i}{\ov{z}_i-\ov{z}_{n+1}}-i\sum\limits_{j=1}^{n-1} \frac{\widetilde{\varepsilon}^{n+1} \cdot k_j}{\ov{z}_n-\ov{z}_{n+1}}+i\frac{(\widetilde{\varepsilon}^{n+1} \cdot q_2) \ov{z}_{n,n+2}}{\ov{z}_{n+2,n+1}\ov{z}_{n,n+1}}\\
     & =i \sum\limits_{i=1}^{n-1}(\widetilde{\varepsilon}^{n+1} \cdot k_i) \frac{\ov{z}_{ni}}{(\ov{z}_i-\ov{z}_{n+1})(\ov{z}_n-\ov{z}_{n+1})}+i\frac{(\widetilde{\varepsilon}^{n+1} \cdot q_2) \ov{z}_{n,n+2}}{\ov{z}_{n+2,n+1}\ov{z}_{n,n+1}}\\
     & =-i \sum\limits_{i=1}^{n-1}(\widetilde{\varepsilon}^{n+1} \cdot k_i) \sum_{l=i}^{n-1} \frac{\ov{z}_{l,l+1}}{(\ov{z}_l-\ov{z}_{n+1})(\ov{z}_{l+1}-\ov{z}_{n+1})} +i\frac{(\widetilde{\varepsilon}^{n+1} \cdot q_2) \ov{z}_{n,n+2}}{\ov{z}_{n+2,n+1}\ov{z}_{n,n+1}}\\
     & =i\Bigg\{ \sum\limits_{l=1}^{n-1}(\widetilde{\varepsilon}^{n+1} \cdot x_l)  \frac{\ov{z}_{l,l+1}}{\ov{z}_{l,n+1}\ov{z}_{n+1,l+1}}+(\widetilde{\varepsilon}^{n+1} \cdot q_2)\frac{\ov{z}_{n+2,n}}{\ov{z}_{n+2,n+1}\ov{z}_{n+1,n}}\Bigg\}=: \Xi^{n+2}\ .
    \end{aligned} \label{NPSI1}
\end{equation}
The above correlator $\Xi^{n+2}$ amounts to the generalized version  of the expression (\ref{cqq}) for the case of more than one graviton. In the last line of \req{NPSI1} the first term in the sum is exactly (\ref{cqq}) and the second term takes into account the additional   interaction with the second graviton. For $\Xi^l$ the upper index $l$ indicates the corresponding pair of Gra\ss mann variables in the anti--holomorphic sector. For the case at hand we have
\begin{equation}
 \langle \theta_{n+2} \ov \theta_{n+2}\;  \widetilde{\varepsilon}^{n+1}_{\mu_1} \ \ov{\partial}\widetilde{X}^{\mu_1}(\ov{z}_{n+1})  \Bigg(\prod\limits_{i=1}^n e^{ik_i \widetilde{X}(\ov{z}_i)}\Bigg)\times e^{iq_{2{\mu_2}}\widetilde{X}^{\mu_2}(\ov{z}_{n+2})} \rangle= \theta_{n+2} \ov \theta_{n+2}\  \Xi^{n+2}\ , \label{n2gg1}
\end{equation}
and similarly for the second graviton we encounter:
\begin{equation}
 \langle  \theta_{n+4} \ov \theta_{n+4}\; \widetilde{\varepsilon}^{n+2}_{\mu_2} \ \ov{\partial}\widetilde{X}^{\mu_2}(\ov{z}_{n+2})  \Bigg(\prod\limits_{i=1}^n e^{ik_i \widetilde{X}(\ov{z}_i)}\Bigg)\times e^{iq_{1{\mu_1}}\widetilde{X}^{\mu_1}(\ov{z}_{n+1})} \rangle= \theta_{n+4} \ov \theta_{n+4}\  \Xi^{n+4}\ .\label{n2gg2}
\end{equation}
Secondly, there is the fermionic piece from the anti--holomorphic parts of the two graviton vertex operators \req{Grav1}:
\begin{align}
     \theta_{n+2}  & \ov{\theta}_{n+2} \theta_{n+4} \ov{\theta}_{n+4}\ \langle \widetilde{\varepsilon}^{n+1}_{\mu_1} \widetilde{\psi}^{\mu_1} q_1 \cdot \widetilde{\psi} \ \widetilde{\varepsilon}^{n+2}_{\mu_2} \widetilde{\psi}^{\mu_2} q_2 \cdot \widetilde{\psi}  \rangle_{S^2}\nonumber\\
     & =\theta_{n+2} \ov{\theta}_{n+2} \theta_{n+4}  \ov{\theta}_{n+4} \Bigg\{-\frac{\widetilde{\varepsilon}_{n+1} \cdot \widetilde{\varepsilon}_{n+2}}{\ov{z}_{n+1}-\ov{z}_{n+2}} \frac{q_1 \cdot q_2}{\ov{z}_{n+1}-\ov{z}_{n+2}}+ \frac{\widetilde{\varepsilon}_{n+1} \cdot q_2}{\ov{z}_{n+1}-\ov{z}_{n+2}} \frac{\widetilde{\varepsilon}_{n+2} \cdot q_1}{\ov{z}_{n+1}-\ov{z}_{n+2}} \Bigg\}\nonumber\\
     &\hskip2cm = -\lf(\ov{\theta}_{n+2} \ov{\theta}_{n+4} \frac{\widetilde{\varepsilon}_{n+1} \cdot \widetilde{\varepsilon}_{n+2}}{\ov{z}_{n+1}-\ov{z}_{n+2}}\ri)\ \lf(\theta_{n+2} \theta_{n+4}  \frac{q_1 \cdot q_2}{\ov{z}_{n+1}-\ov{z}_{n+2}}\ri)\nonumber\\
     &\hskip2cm + \lf( \ov{\theta}_{n+2} \theta_{n+4}\frac{\widetilde{\varepsilon}_{n+1} \cdot q_2}{\ov{z}_{n+1}-\ov{z}_{n+2}}\ri)\ \lf(\theta_{n+2} \ov{\theta}_{n+4}  \frac{\widetilde{\varepsilon}_{n+2} \cdot q_1}{\ov{z}_{n+1}-\ov{z}_{n+2}}\ri)\ .\label{n2gg3}
    \end{align}
The expressions (\ref{n2gg1}), (\ref{n2gg2}) and (\ref{n2gg3}) can be extracted from the last four terms of the exponential (\ref{Gen2}). 
Let us now comprise  all terms into a matrix. We have the following relation 
\begin{equation}
    \begin{aligned}
     \Big\langle  \Bigg(\prod\limits_{i=1}^n e^{ik_i \widetilde{X}(\ov{z}_i)}\Bigg)&  \exp\{i \widetilde{q}_1\cdot \widetilde{X}+\theta_{n+2}\bar{\theta}_{n+2} \widetilde{\varepsilon}_{n+1} \cdot \ov{\partial}\widetilde{X} +\theta_{n+2} \widetilde{q}_1\cdot \widetilde{\psi}+\bar{\theta}_{n+2} \widetilde{\varepsilon}_{n+1} \cdot \widetilde{\psi}  \} \\
    & \times   \exp\{i \widetilde{q}_2\cdot \widetilde{X}+\theta_{n+4}\bar{\theta}_{n+4} \widetilde{\varepsilon}_{n+2} \cdot \ov{\partial}\widetilde{X} +\theta_{n+4} \widetilde{q}_2\cdot \widetilde{\psi}+\bar{\theta}_{n+4} \widetilde{\varepsilon}_{n+2} \cdot \widetilde{\psi} \} \Big\rangle_{S_2}\Bigg|_{\alpha' \rightarrow \infty}\\
    &  = \exp \Bigg\{ \frac{1}{2}\alpha'^2\sum\limits_{i,j \in\{n+2,n+4\}} \begin{pmatrix} \theta_i \\ \bar{\theta}_i  \end{pmatrix}^t \Psi_{2}\,\, \begin{pmatrix} \theta_j \\ \bar{\theta}_j  \end{pmatrix} \Bigg\}+\mathcal{O}(\ap^{-1})\ ,
    \end{aligned}
\end{equation}
where $\Psi_{2}$ is given as:
\begin{equation}
    \begin{aligned}
     & \Psi_{2}=\begin{pmatrix} 
    0 & \frac{q_1 \cdot q_2}{\ov{z}_{n+1,n+2}} & \Xi^{n+2} & \frac{\widetilde{\varepsilon}_{n+2} \cdot q_1}{\ov{z}_{n+2,n+1}} 
 \\   -\frac{q_1 \cdot q_2}{\ov{z}_{n+1,n+2}} & 0 & \frac{-\widetilde{\varepsilon}_{n+1} \cdot q_2}{\ov{z}_{n+1,n+2}}  & \Xi^{n+4}
 \\   -\Xi^{n+2} &  \frac{\widetilde{\varepsilon}_{n+1} \cdot q_2}{\ov{z}_{n+1,n+2}}  & 0 &  \frac{\widetilde{\varepsilon}_{n+1} \cdot \widetilde{\varepsilon}_{n+2}}{\ov{z}_{n+1,n+2}} 
 \\  - \frac{\widetilde{\varepsilon}_{n+2} \cdot q_1}{\ov{z}_{n+2,n+1}} & -\Xi^{n+4} & -\frac{\widetilde{\varepsilon}_{n+1} \cdot \widetilde{\varepsilon}_{n+2}}{\ov{z}_{n+1,n+2}}  & 0
 \end{pmatrix}\ .
    \end{aligned} \label{show}
\end{equation}
The matrix \req{show} is the same expression (denoted by $\Psi_S$) defined in eq. (3.8) of \cite{chy} for the two--graviton case. 
Here, $\Xi^{n+2}$ and $\Xi^{n+4}$ are the objects  introduced in (\ref{n2gg1}) and (\ref{n2gg2}) and correspond to the two anti--holomorphic graviton fields labelled by $n+2$ and $n+4$, respectively. 
Putting everything together, for the anti holomorphic part we have:
\begin{equation}
\begin{aligned}
\overline{KN}^{-1}\ \mathcal{I}(n;2)&\Big|_{\substack{{\mbox{\tiny anti-}\atop\mbox{\tiny  holomorphic}}\\ \alpha' \rightarrow \infty}}= \mathcal{C}(1,2,\ldots,n)\\
&\times\int \prod\limits_{i\in \{n+2,n+4\}} d\theta_i d\bar\theta_i\ \exp \Bigg\{\h\alpha'^2 \sum\limits_{i,j \in \{n+2,n+4\}}  \begin{pmatrix} \theta_i \\ \bar{\theta}_i  \end{pmatrix}^t \Psi_{2}\,\, \begin{pmatrix} \theta_j \\ \bar{\theta}_j  \end{pmatrix}\Bigg\}+\mathcal{O}(\ap^{-1}) \\[2mm]
& = \mathcal{C}(1,2,\ldots,n)\  \pf\Psi_{2}+\mathcal{O}(\ap^{-1})\ .
\end{aligned}\label{ANTIH}
\end{equation} 
As a second for the holomorphic part of \req{NPSI1} we are dealing with  $n+r=n+2$ open superstring vertex operators which can be described by the pure open superstring disk amplitude, which is related to the Pfaffian of the  matrix $\Psi_{n+2}$:
    \begin{equation}
    \begin{aligned}
 KN^{-1}\    \mathcal{I}(n;2)\Big|_{\substack{\mbox{\tiny  holomorphic}\\ \alpha' \rightarrow \infty}}& =\Bigg\langle \prod\limits_{j=1}^n V_o(\varepsilon_j,k_j,z_j) \prod\limits_{l=1}^2 V_o(\varepsilon_l,q_l,z_l)\Bigg\rangle_{S^2}\Big|_{ \alpha' \rightarrow \infty} \\
    & =\Bigg\langle \prod\limits_{j=1}^n V_o(\varepsilon_j,k_j,z_j) \prod\limits_{l=1}^2 V_o(\varepsilon_l,q_l,z_l)\Bigg\rangle_{D^2}\Big|_{ \alpha' \rightarrow \infty} =\pf' \Psi_{n+2}\ .
    \end{aligned}   \label{PYM}
\end{equation} 
It is clear that these contractions  give rise to the first four terms of the integrand (\ref{Gen2}) as they are identical to contractions of the $n+2$ open strings given in the standard formula (\ref{twistedWitten}). 
After putting together the  anti--holomorphic \req{ANTIH} and holomorphic \req{PYM} parts   we have
\begin{equation}
\begin{aligned}
 \lim_{\ap\ra\infty}\ 
\overline{KN}^{-1}\  \mathcal{I}(n;2)\Big|_{{\mbox{\tiny anti-}\atop\mbox{\tiny  holomorphic}}}\cdot  KN^{-1}\  \mathcal{I}(n;2)\Big|_{\mbox{\tiny  holomorphic}}&= \mathcal{C}(1,2,\ldots,n)\\ 
&\times\pf \Psi_{2} \  \pf' \Psi_{n+2}+\mathcal{O}(\ap^{-1})\ , \label{n+2}
\end{aligned}
\end{equation}
with the $(2n+4)\times(2n+4)$ matrix  $\Psi_{n+2}$ and the $4\times4$ matrix $\Psi_2$ both assuming the form of \req{PSI} with entries (\ref{psi2})
\begin{align}
\Psi_{n+2}&=\Psi_S=\lf.\begin{pmatrix} A_{ij} & -C_{ji} \\ C_{ij} & B_{ij} \end{pmatrix}\ri|_{\sigma_l=\zeta_l,\ l=1,\ldots,n+1\atop{\sigma_{n+2}=\zeta_{n+3}}}\ \ ,\ \ S=\{1,\ldots,n,n+1,n+2\}\ ,\label{psin2}\\[3mm]
\Psi_2&=\Psi_{S_2}=\lf. \begin{pmatrix} A_{ij} & -C_{ji} \\ C_{ij} & B_{ij} \end{pmatrix}\ri|_{\sigma_l=\bar\zeta_l,\ l=1,\ldots,n\atop{\sigma_{n+1}=\bar\zeta_{n+2},\ \sigma_{n+2}=\bar\zeta_{n+4}}}\ \ ,\ \ S_2=\{n+1,n+2\}\ ,\label{psi12}
\end{align}
respectively. Finally,  noting the expansion of denominator terms w.r.t. $\ap$
\be\label{Noting}
\frac{\ov{\theta}_i \ov{\theta}_j (\xi_i \cdot \xi_j)}{\zeta_i-\zeta_j\mp \ap^{-1} \theta_i \theta_j}=\frac{\ov{\theta}_i \ov{\theta}_j (\xi_i \cdot \xi_j)}{\zeta_i-\zeta_j}\pm \ap^{-1}\frac{\theta_i \theta_j \ov{\theta}_i \ov{\theta}_j (\xi_i \cdot \xi_j)}{(\zeta_i-\zeta_j)^2}\ ,
\ee
with \req{psin2} and \req{psi12} we can write the integrand \req{Gen2} in the terms of the generalized notation \req{defsetss} as the following product:
\begin{equation}{\small
    \begin{aligned}
    \mathcal{I}(n;2) &=\int  \prod\limits_{i=1}^{n+4} d \theta_i d\bar{\theta_i} \ \frac{\theta_1 \theta_2}{\zeta_1-\zeta_2} \exp\Bigg\{\frac{1}{2}\ap^2\sum\limits_{ i,j=1\atop i,j \neq n+2}^{n+3} \begin{pmatrix} \theta_i \\ \bar{\theta}_i  \end{pmatrix}^t  \Psi_{n+2} \,\, \begin{pmatrix} \theta_j \\ \bar{\theta}_j  \end{pmatrix}\pm \h\ap\sum\limits_{ i,j=1\atop i,j \neq n+2}^{n+3}\frac{\theta_i \theta_j \ov{\theta}_i \ov{\theta}_j (\xi_i \cdot \xi_j)}{(\zeta_i-\zeta_j)^2}\Bigg\}\\
 &  \times \exp\Bigg\{\frac{1}{2}\ap^2\sum\limits_{i,j\in \{n+2,n+4\}} \begin{pmatrix} \theta_i \\ \bar{\theta}_i  \end{pmatrix}^t \Psi_2 \,\, \begin{pmatrix} \theta_j \\ \bar{\theta}_j  \end{pmatrix} \pm \h\ap\sum\limits_{i,j\in \{n+2,n+4\}}\frac{\theta_i \theta_j \ov{\theta}_i \ov{\theta}_j (\xi_i \cdot \xi_j)}{(\ov{\zeta}_i-\ov{\zeta}_j)^2}\Bigg\}\\
 &\times  \mathcal{C}(1,2,\ldots,n)\ \times KN \cdot \overline{KN}.
  \end{aligned} } \label{n2ggtw}
\end{equation} 
Our integrand \req{n2ggtw} furnishes a KLT like structure factorizing
holomorphic~and anti--holomorphic terms. In addition, to this factorized form the  two sets of fermionic variables $\bigcup\limits_{j\in\{1,\ldots,n+1,n+3\}}\{\theta_j,\bar\theta_j\}$ and $\bigcup\limits_{i\in\{n+2,n+4\}}\{\theta_i,\bar\theta_i\}$  can be attributed, respectively.

Eventually, after these preparations we are able to  construct the pair of twisted  forms $\varphi^{EYM}_{\pm,n;2}$ and $\widetilde{\varphi}^{EYM}_{\pm,n;2}$.  Similarly to \req{stcnew} we take the pure holomorphic part of \req{n2ggtw}   to constitute the twisted form $\varphi^{EYM}_{\pm,n;2}$
\begin{align}
 \varphi^{EYM}_{\pm,n;2}&=  d\mu_{n+2}\ \int  \prod\limits_{ i=1\atop i \neq n+2}^{n+3} d \theta_i d\bar{\theta_i}\ \frac{\theta_1 \theta_2}{\zeta_1-\zeta_2}\  \exp\Bigg\{\frac{1}{2}\ap^2\sum\limits_{ i,j=1\atop i,j\neq n+2}^{n+3} \begin{pmatrix} \theta_i \\ \bar{\theta}_i  \end{pmatrix}^t \Psi_{n+2} \,\, \begin{pmatrix} \theta_j \\ \bar{\theta}_j  \end{pmatrix}\Bigg\}\nonumber\\  
&\times   \exp\Bigg\{\pm \frac{1}{2} \ap \sum\limits_{ i,j=1\atop i,j \neq n+2}^{n+3} \frac{\theta_i \theta_j \ov{\theta}_i \ov{\theta}_j (\xi_i \cdot \xi_j)}{(\zeta_i-\zeta_j)^2}  \Bigg\}\ ,
 \end{align} 
with $\Psi_{n+2}$ defined in \req{psin2}. To find our second twisted form $\widetilde{\varphi}^{EYM}_{\pm,n;2}$ we take the anti--holomorphic part of \req{n2ggtw} and apply the isomorphism \req{ISOMAP} to arrive at:
\begin{align}
   \widetilde{\varphi}^{EYM}_{\pm,n;2}&= d\mu_{n+2} \  \mathcal{C}(1,2,\ldots,n) \ \int   \prod\limits_{ i \in \{n+2,n+4\}} d \theta_i d\bar{\theta_i}\ 
   \exp\Bigg\{\frac{1}{2}\ap^2\sum\limits_{i,j\in \{n+2,n+4\}} \begin{pmatrix} \theta_i \\ \bar{\theta}_i  \end{pmatrix}^t \Psi_2 \,\, \begin{pmatrix} \theta_j \\ \bar{\theta}_j  \end{pmatrix}\Bigg\} \Bigg|_{\ov{\zeta}_l \rightarrow \zeta_l}\nonumber\\
  &\times   \exp\Bigg\{\pm \frac{1}{2} \ap\sum\limits_{i,j\in \{n+2,n+4\}}\frac{\theta_i \theta_j \ov{\theta}_i \ov{\theta}_j (\xi_i \cdot \xi_j)}{({\zeta}_i-{\zeta}_j)^2}  \Bigg\}\ .\end{align}
Our two twisted forms $\varphi_-=\widetilde{\varphi}^{EYM}_{+,n;2}$
and $\varphi_+= \varphi^{EYM}_{-,n;2}$ reproduce the EYM integrand in the $\ap\ra\infty$ limit
\req{expansion}.
These limits can be determined as:
 \begin{equation}
    \begin{aligned}
 \lim_{\ap \rightarrow \infty}\hat{\varphi}^{EYM}_{\pm,n;2}&= \int  \prod\limits_{ i=1\atop i \neq n+2}^{n+3} d \theta_i d\bar{\theta_i}\ \frac{\theta_1 \theta_2}{\zeta_1-\zeta_2}\ \exp\Bigg\{\frac{1}{2}\ap^2\sum\limits_{ i,j=1\atop i,j\neq n+2}^{n+3} \begin{pmatrix} \theta_i \\ \bar{\theta}_i  \end{pmatrix}^t \Psi_{n+2}\,\, \begin{pmatrix} \theta_j \\ \bar{\theta}_j  \end{pmatrix}\Bigg\}\pm \mathcal{O}(\ap^{-1}) \\[2mm]
  & =\frac{\pf \Psi_{n+2}^{12}}{\zeta_1-\zeta_2}+\mathcal{O}(\ap^{-1})= \pf' \Psi_{n+2}+\mathcal{O}(\ap^{-1})\ ,\\
  & \\
   \lim_{\ap \rightarrow \infty}\hat{\widetilde{\varphi}}^{EYM}_{\pm,n;2}&= \mathcal{C}(1,2,\ldots,n) \ \int  \prod\limits_{ i \in\{n+2,n+4\}} d \theta_i d\bar{\theta_i}  
  \exp\Bigg\{\frac{1}{2}\ap^2\sum\limits_{i,j\in \{n+2,n+4\}} \begin{pmatrix} \theta_i \\ \bar{\theta}_i  \end{pmatrix}^t \Psi_2 \,\, \begin{pmatrix} \theta_j \\ \bar{\theta}_j  \end{pmatrix}\Bigg\}\Bigg|_{\ov{\zeta}_l \rightarrow \zeta_l}\\[2mm]
  &\pm \mathcal{O}(\ap^{-1})  =\mathcal{C}(1,2,\ldots,n) \ \pf \Psi_2\big|_{\ov{\zeta}_l \rightarrow \zeta_l}+\mathcal{O}(\ap^{-1})
\end{aligned}
\end{equation}
Using these limits in  (\ref{expansion}) gives the final result for EYM amplitude  in the CHY formalism: 
\begin{equation}
    \begin{aligned}
       \mathcal{A}(n;2)&=\lim_{\ap \rightarrow \infty} \langle \widetilde{\varphi}^{EYM}_{+,n;2}, \varphi_{-,n;2}^{EYM}\rangle_\om
       =\int\limits_{\mathcal{M}_{0,n+2}}\!\!\! d \mu_{n+2}\ 
 \sideset{}{'}\prod_{a=1}^{n+2} \delta(f_a) \lim_{\alpha' \rightarrow \infty} 
  \hat{\varphi}_{-,n;2}^{EYM}\ \hat{ \widetilde{\varphi}}^{EYM}_{+,n;2}\\
 &=\!\!\!\!\int\limits_{\mathcal{M}_{0,n+2}}\!\!\! d \mu_{n+2}
 \sideset{}{'}\prod_{a=1}^{n+2} \delta(f_a)\;  \frac{\pf \Psi_2\big|_{\ov{\zeta}_l\ \rightarrow \zeta_l}\ \pf' \Psi_{n+2}}{(z_1-z_2)(z_2-z_3)\ldots(z_n-z_1)}\ \ .
       \end{aligned}
\end{equation}

\subsection{Twisted form and intersections for amplitudes of $n$ gluons and $r$ gravitons}

Finally, in this subsection we extend our results for EYM amplitudes  to the all multiplicity case. We shall consider the EYM amplitude involving $n$ gluons and $r$ gravitons.
 We start with the correlator \req{Startn1} on the sphere, which can be expressed in terms of Gra\ss mann variables as given in (\ref{IN1}).
Performing the Wick contractions on the sphere we obtain for the integrand:
\begin{equation}
    \begin{aligned}
 \mathcal{I}(n;r)  & = \int   \prod\limits_{i=1}^{n+2r}  \frac{\theta_1 \theta_2}{\zeta_1-\zeta_2}\ d \theta_i d\bar{\theta_i}\  \exp\Bigg\{ \alpha'^2 \Bigg( \sum\limits_{i,j,\in\Sc \atop i\neq j}\frac{ (\bar{\theta}_i \theta_j \xi_i \cdot p_j)}{\zeta_i-\zeta_j \mp \alpha'^{-1} \theta_i \theta_j}+\sum\limits_{i,j,\in\Sc \atop i\neq j}\frac{ (\theta_i \bar{\theta_i} \xi_i \cdot p_j)}{\zeta_i-\zeta_j \mp \alpha'^{-1} \theta_i \theta_j}\\
&    +\sum\limits_{i,j,\in\Sc\atop i>j} \frac{( \bar{\theta_i} \bar{\theta_j}\xi_i \cdot \xi_j)}{\zeta_i-\zeta_j \mp \alpha'^{-1} \theta_i \theta_j}+\sum\limits_{i,j,\in\Sc\atop i>j}\frac{( \theta_i \theta_j p_i \cdot p_j)}{\zeta_i-\zeta_j \mp \alpha'^{-1} \theta_i \theta_j}+\sum\limits_{j \in \{1,...,n\} \cup \Sc_r \atop i\in \Sc_r}\frac{ (\theta_i \bar{\theta_i} \xi_i \cdot p_j)}{\ov{\zeta}_i-\ov{\zeta}_j \mp \alpha'^{-1} \theta_i \theta_j}\\
&+  \sum\limits_{i,j\in\Sc_r \atop i\neq j}\frac{ (\bar{\theta}_i \theta_j \xi_i  \cdot p_j)}{\ov{\zeta}_i-\ov{\zeta}_j \mp \alpha'^{-1} \theta_i \theta_j}+\sum\limits_{i,j\in\Sc_r\atop i>j} \frac{( \bar{\theta_i} \bar{\theta_j}\xi_i  \cdot \xi_j)}{\ov{\zeta}_i-\ov{\zeta}_j \mp \alpha'^{-1} \theta_i \theta_j}+\sum\limits_{i,j\in\Sc_r\atop i>j}\frac{( \theta_i \theta_j p_i \cdot p_j)}{\ov{\zeta}_i-\ov{\zeta}_j \mp \alpha'^{-1} \theta_i \theta_j} \Bigg)\Bigg\}\\[2mm]
& \times \mathcal{C}(1,2,\ldots,n)\times KN \cdot \ov{KN}\ .\label{genint}
     \end{aligned}
\end{equation} 
In \req{genint} we have introduced the two sets $\Sc$ and $\Sc_r$, given by:
\begin{equation}
\begin{aligned}
& \Sc:=\{1,2,3,...,n,n+1,n+3,...,n+2r-1\}\ ,\\
& \Sc_r:=\{n+2,n+4,...,n+2r\}\ .
\end{aligned}
\end{equation}
Here, $\Sc_r$  represents the set  of indices accounting for the anti--holomorphic parts of the $r$ graviton vertex operators \req{Grav1}, while  $\Sc$ is the set of indices labelling  the holomorphic parts of both gluons and gravitons.
Furthermore,  in the exponential of \req{genint} the sums run over indices denoting  fermionic variables $\th_i,\ov\th_i$, the set of the generalized momenta $p_i$ given in \req{P}, polarizations $\xi_j$ defined in \req{pol} and positions $\zeta_j$  defined  (\ref{zeta}), respectively. 
As in  \req{Gen2} the first four terms in the exponential account for the holomorphic
field contractions, while the last four terms represent the anti--holomorphic field contractions subject to the map \req{MAP}.
Furthermore, $KN\cdot \overline{KN}$ is the Koba--Nielsen factor  \req{KN} for $n+r$ closed strings.

We shall follow  similar steps as in the previous subsection to extract from \req{genint} a pair of twisted forms suitable for describing the multi--leg EYM amplitude.
Let us first look at the  CHY integral introduced in (\ref{AmpEYM}) for this general case 
\begin{equation}
    \mathcal{I}_{n+r}(n;r)=\mathcal{C}(1,2,\ldots,n)\ \pf \Psi_{r}\ \pf'\Psi_{n+r}(k_a,q_a,\varepsilon,\sigma)\ , \label{GENCHY}
\end{equation}
with $\mathcal{C}(1,2,\ldots,n)$ being the Parke--Taylor factor \req{ColorForm}.
As in our constructions above the latter is to be identified with the gauge current factor $\mathcal{C}$ in \req{genint} which  enters the definition  of  $ \widetilde{\varphi}^{EYM}_{\pm,n;r}$. 

Again, by using \req{Noting} in the exponential of the integrand \req{genint}  we may disentangle quadratic from linear orders in $\ap$. As a consequence 
the eight sums accounting for the quadratic order $\ap^2$  can compactly be written as
\begin{equation}
\sum\limits_{ i,j\in \Sc} ( \theta_i \,\, \bar{\theta}_i ) \Psi_{n+r} \,\, \begin{pmatrix} \theta_j \\ \bar{\theta}_j  \end{pmatrix}+\sum\limits_{i,j\in\Sc_r}  ( \theta_i \,\, \bar{\theta}_i ) \Psi_{r} \,\, \begin{pmatrix} \theta_j \\ \bar{\theta}_j  \end{pmatrix}\ , \label{NNEsted}
\end{equation}
with the $(2n+2r)\times(2n+2r)$ matrix  $\Psi_{n+r}$ and the $2r\times2r$ matrix $\Psi_r$ both assuming the form of \req{PSI} with entries (\ref{psi2})
\begin{align}
\Psi_{n+r}&=\Psi_S=\lf.\begin{pmatrix} A_{ij} & -C_{ji} \\ C_{ij} & B_{ij} \end{pmatrix}\ri|_{\sigma_l=\zeta_l,\ l=1,\ldots,n\atop{\sigma_{n+k}=\zeta_{n+2k-1},\; k=1,\ldots,r}}\ ,\ \ S=\{1,\ldots,n,n+1,\ldots,n+r\}\ ,\label{Psin2}\\[3mm]
\Psi_r&=\Psi_{S_r}=\lf. \begin{pmatrix} A_{ij} & -C_{ji} \\ C_{ij} & B_{ij} \end{pmatrix}\ri|_{\sigma_l=\bar\zeta_l,\ l=1,\ldots,n\atop{\sigma_{n+k}=\bar\zeta_{n+2k},\; k=1,\ldots,r}}\ ,\ \ S_r=\{n+2,\ldots,n+2r\}\ ,\label{Psi12}
\end{align}
respectively.
After taking into account the linear $\ap$ order originating from the expansion \req{Noting} with the above matrices \req{Psin2} and \req{Psi12} the integrand \req{genint} can be cast into:
\begin{equation}
    \begin{aligned}
   \mathcal{I}(n;r)&=  \int  \prod\limits_{i=1}^{n+2r} d \theta_i d\bar{\theta_i} \ \frac{\theta_1 \theta_2}{\zeta_1-\zeta_2}\  \exp\Bigg\{\frac{1}{2}\ap^2\sum\limits_{ i,j\in\Sc} \begin{pmatrix} \theta_i \\ \bar{\theta}_i  \end{pmatrix}^t \Psi_{n+r} \begin{pmatrix} \theta_j \\ \bar{\theta}_j  \end{pmatrix}\pm \h\ap\sum\limits_{ i,j\in\Sc}\frac{\theta_i \theta_j \ov{\theta}_i \ov{\theta}_j (\xi_i \cdot \xi_j)}{(\zeta_i-\zeta_j)^2}\Bigg\}\\[2mm]
 &  \times  \exp\Bigg\{\frac{1}{2}\ap^2\sum\limits_{i,j\in\Sc_r} \lf(\th_i\atop\bar\th_i\ri)^t\Psi_r \,\, \begin{pmatrix} \theta_j \\ \bar{\theta}_j  \end{pmatrix} \pm \h\ap\sum_{i,j\in\Sc_r}\frac{\theta_i \theta_j \ov{\theta}_i \ov{\theta}_j (\xi_i \cdot \xi_j)}{(\ov{\zeta}_i-\ov{\zeta}_j)^2}\Bigg\}\\[1mm]
 & \times  \mathcal{C}(1,2,\ldots,n)\times KN\cdot\overline{KN}\ .
 \end{aligned} \label{INTEGNN}
\end{equation}
Again, our integrand \req{INTEGNN} furnishes a KLT like structure factorizing
holomorphic~and anti--holomorphic terms. In addition, to this factorized form in lines of \req{NNEsted} the  two sets of fermionic variables $\bigcup\limits_{j\in\Sc}\{\theta_j,\bar\theta_j\}$ and $\bigcup\limits_{i\in\Sc_r}\{\theta_i,\bar\theta_i\}$  can be attributed, respectively.

Now, we are prepared  to construct the pair of twisted forms $\varphi^{EYM}_{\pm,n;r}$ and $\tilde\varphi^{EYM}_{\pm,n;r}$. As  first differential form we define:
\begin{align}
   \varphi^{EYM}_{\pm,n;r}&=  d\mu_{n+r}\ \int  \prod\limits_{ i\in\Sc} \frac{\theta_1 \theta_2}{\zeta_1-\zeta_2}\ d \theta_i 
   d\bar{\theta_i}\label{nrtwistedgen1}\\  
   &\times\exp\Bigg\{\frac{1}{2} \ap^2\sum\limits_{i,j\in\Sc} \begin{pmatrix} \theta_i \\ \bar{\theta}_i  \end{pmatrix}^t \Psi_{n+r} \,\, \begin{pmatrix} \theta_j \\ \bar{\theta}_j  \end{pmatrix}\Bigg\} \exp\Bigg\{\pm \frac{1}{2} \ap \sum\limits_{i,j\in\Sc} \frac{\theta_i \theta_j \ov{\theta}_i \ov{\theta}_j (\xi_i \cdot \xi_j)}{(\zeta_i-\zeta_j)^2}  \Bigg\}\ .\nonumber
    \end{align} 
In this form $\varphi^{EYM}_{\pm,n;r}$ can be identified with the twisted gauge form \req{Witten}, 
i.e.: 
\be\label{Subject}
\varphi^{EYM}_{\pm,n;r}\equiv \varphi^{gauge}_{\pm,n+r}\;\Bigg|^{z_l=\zeta_l,\ l=1,\ldots,n\atop{z_{n+k}=\zeta_{n+2k-1},\; k=1,\ldots,r}}_{\theta_{n+k}=\theta_{n+2k-1}\  \ \ \  \ \ \ \ \ \  \atop \bar\theta_{n+k}=\bar \theta_{n+2k-1},\ k=1,\ldots,r\ .}
\ee    
For the second twisted form  $\widetilde{\varphi}^{EYM}_{\pm,n;r}$ we take the 
anti--holomorphic part of (\ref{genint}) and  apply the isomorphism \req{ISOMAP}:
   \begin{align}
     \widetilde{\varphi}^{EYM}_{\pm,n;r}&=  d\mu_{n+r}\ \mathcal{C}(1,2,\ldots,n)\ \int  \prod\limits_{ i\in\Sc_r}   d \theta_i d\bar{\theta_i} \ \exp\Bigg\{\frac{1}{2} \ap^2\sum\limits_{i,j\in \Sc_r} \begin{pmatrix} \theta_i \\ \bar{\theta}_i  \end{pmatrix}^t \Psi_r  \,\, \begin{pmatrix} \theta_j \\ \bar{\theta}_j  \end{pmatrix}\Bigg\}\Bigg|_{\ov{\zeta_l} \rightarrow \zeta_l}\nonumber\\[2mm]
   &  \times \exp\Bigg\{\pm \frac{1}{2} \ap \sum\limits_{i,j\in \Sc_r} \frac{\theta_i \theta_j \ov{\theta}_i \ov{\theta}_j (\xi_i \cdot \xi_j)}{(\zeta_i-\zeta_j)^2}  \Bigg\} \ .\label{nrtwistedgen2}
    \end{align}
With these two twisted forms \req{nrtwistedgen1} and \req{nrtwistedgen2} we can compute the EYM amplitude through the expression for the intersection number \req{expansion} in the $\ap \rightarrow \infty$ limit. For this we determine the following limits:
\begin{align}
  \lim_{\ap \rightarrow \infty}\hat{\varphi}^{EYM}_{\pm,n;r}& = \int  \prod\limits_{ i\in\Sc_r}   d \theta_i d\bar{\theta_i}\frac{\theta_1 \theta_2}{\zeta_1-\zeta_2} \exp\Bigg\{\frac{1}{2} \ap^2\sum\limits_{i,j\in\Sc} \begin{pmatrix} \theta_i
  \\ \bar{\theta}_i  \end{pmatrix}^t \Psi_{n+r} \,\, \begin{pmatrix} \theta_j \\ \bar{\theta}_j  \end{pmatrix}\Bigg\} +\mathcal{O}(\ap^{-1})  \nonumber\\
   & = \frac{\pf \Psi_{n+r}^{12}}{\zeta_1-\zeta_2}+\mathcal{O}(\ap^{-1})= \pf' \Psi_{n+r}+\mathcal{O}(\ap^{-1})\ ,\nonumber\\[3.5mm]
    \lim_{\ap \rightarrow \infty}\hat{\widetilde{\varphi}}^{EYM}_{\pm,n;r}& = \mathcal{C}(1,2,\ldots,n)\ \int  \prod\limits_{i\in\Sc_r}   d \theta_i d\bar{\theta_i}\  \exp\Bigg\{\frac{1}{2} \ap^2\sum\limits_{i,j\in\Sc_r}  \begin{pmatrix} \theta_i \\ \bar{\theta}_i  \end{pmatrix}^t \Psi_r  \,\, \begin{pmatrix} \theta_j \\ \bar{\theta}_j  \end{pmatrix}\Bigg\}\Bigg|_{\ov{\zeta}_l \rightarrow \zeta_l} +\mathcal{O}(\ap^{-1}) \nonumber\\
   & =\mathcal{C}(1,2,\ldots,n)\ \pf \Psi_r\big|_{\ov{\zeta}_l \rightarrow \zeta_l}+\mathcal{O}(\ap^{-1})\ .\label{Limits} 
    \end{align}
Putting the limits \req{Limits} into (\ref{expansion}) subject to the choice $\varphi_+= \widetilde{\varphi}^{EYM}_{+,n;r}$ and $\varphi_- =\varphi^{EYM}_{-,n;r}$ yields the EYM amplitude for $n$ gluons and $r$ gravitons:
\begin{equation}
    \begin{aligned}
       \mathcal{A}(n;r)&=\lim_{\ap \rightarrow \infty} \langle \widetilde{\varphi}^{EYM}_{+,n;r}, \varphi_{-,n;r}^{EYM}\rangle_\om
       =\int\limits_{\mathcal{M}_{0,n+r}}\!\!\! d \mu_{n+r}\ 
 \sideset{}{'}\prod_{a=1}^{n+r} \delta(f_a) \lim_{\alpha' \rightarrow \infty} 
  \hat{\varphi}_{-,n;r}^{EYM}\ \hat{ \widetilde{\varphi}}^{EYM}_{+,n;r}\\
 &=\!\!\!\!\int\limits_{\mathcal{M}_{0,n+r}}\!\!\! d \mu_{n+r}
 \sideset{}{'}\prod_{a=1}^{n+r} \delta(f_a)\;  \frac{\pf \Psi_r\big|_{\ov{\zeta}_l \rightarrow \zeta_l}\ \pf' \Psi_{n+r}}{(z_1-z_2)(z_2-z_3)\ldots(z_n-z_1)}\ \ .
       \end{aligned}\label{FinalAmp}
\end{equation}
This is the integral formula for the Einstein Yang-Mills amplitude \req{AmpEYM} formulated in the CHY formalism for the generic case of $n$ gluons and $r$ gravitons. 

\subsection{Decomposing EYM amplitudes in terms of  pure gluon subamplitudes}

Finally, in view of the results \cite{Stieberger:2014cea,Stieberger:2015qja,Stieberger:2015kia,Stieberger:2015vya,ST} where EYM amplitudes are expressed in terms of pure gluon amplitudes in this subsection we shall use our pair of twisted forms 
 \req{nrtwistedgen1} and \req{nrtwistedgen2} and work out some decomposition subject to twisted intersection theory.
With $m=n+r$ and following \cite{MT}  we may expand $\widetilde{\varphi}^{EYM}_{+,n;r}$ w.r.t. to a orthonormal basis  
 of $(n+r)$--forms $\bigcup\limits_{a=1}^{(m-3)!}\{\Phi_{+,a}\}\in H_{+\om}^{m-3}$ and 
 $\varphi_{-,n;r}^{EYM}$ w.r.t. to its dual basis $\bigcup\limits_{b=1}^{(m-3)!}\{\Phi^\vee_{-,b}\}\in H_{-\om}^{m-3}$ as
 \begin{align}
 \widetilde{\varphi}^{EYM}_{+,n;r}&=\sum_{a=1}^{(m-3)!}\vev{\Phi^\vee_{-,a}, \widetilde{\varphi}^{EYM}_{+,n;r} }_\om \ \Phi_{+,a}\ ,\\
 \varphi_{-,n;r}^{EYM}&=\sum_{b=1}^{(m-3)!}\vev{\varphi_{-,n;r}^{EYM}   , \Phi_{+,b}}_\omega\ \Phi^\vee_{-,b}\ ,
\end{align}
respectively.
With the intersection matrix $\vev{\Phi_{+,a},\Phi^\vee_{-,b}}_\om=\delta_{ab}$ we then can write 
our  EYM amplitude \req{FinalAmp} as:
\begin{align}
       \mathcal{A}(n;r)&=\lim_{\ap \rightarrow \infty} \langle \widetilde{\varphi}^{EYM}_{+,n;r}, \varphi_{-,n;r}^{EYM}\rangle_\om\label{ExTwisted}\\
&=\lim_{\ap \rightarrow \infty} 
  \sum_{a,b=1}^{(m-3)!}
    \vev{\Phi^\vee_{-,a}, \widetilde{\varphi}^{EYM}_{+,n;r} }_\om \ \delta_{ab} \ 
       \vev{\varphi_{-,n;r}^{EYM}   , \Phi_{+,b}  }_\omega\ .\nonumber
 \end{align}
 If our basis is represented by Parke--Taylor forms \req{PT}, i.e. $\Phi_{+,b}=PT(b)$
 with \req{Subject} the intersection number $\vev{\varphi_{-,n;r}^{EYM}   , \Phi_{+,b}  }_\omega$ becomes the
 $(n+r)$ gluon subamplitude \req{gaugeinter} of ordering $b$  being independent on $\ap$, i.e.
 $\vev{\varphi_{-,n;r}^{EYM}   , \Phi_{+,b}  }_\omega=\Ac_{YM}(b)$.  The dual Parke--Taylor forms $PT^\vee(a)$ have been introduced in \cite{MT}
  \begin{align}\label{dualPT}
 PT^\vee(a)&\equiv PT^\vee(1,a(2),\ldots,a(m-2),m-1,m)\\
 &=\fc{d\mu_m}{(\zeta_1-\zeta_{m-1})(\zeta_{m-1}-\zeta_m )(\zeta_1-\zeta_m)} \prod_{i=2}^{m-2}\sum_{j=1}^{i-1}\fc{(p_{a(i)}+p_{a(j)})^2}{\z_{a(i)}-\z_{a(j)}}\ 
 \fc{\z_{a(j)}-\z_{m}}{\z_{a(i)}-\z_{m}}\ ,\nonumber
 \end{align}
 and are related to the integrands of the $m$ open superstring amplitude  \cite{Mafra:2011nv,Mafra:2011nw}.
 To this end \req{ExTwisted} becomes
 \be\label{result}
\mathcal{A}(n;r)=\lim_{\ap \rightarrow \infty} \
  \sum_{a=1}^{(m-3)!}
    \vev{PT^\vee(a), \widetilde{\varphi}^{EYM}_{+,n;r} }_\om \ \ \Ac_{YM}(a)\ ,
 \ee
 which expresses any EYM amplitude $\mathcal{A}(n;r)$ in terms of a linear combination of a basis of 
 $(m-3)!$  pure $m$ gluon subamplitudes $\Ac_{YM}(a)$.
 Note, that in \req{result} only the twisted form $\widetilde{\varphi}^{EYM}_{+,n;r}$ depends on $\ap$. Consequently, the limit $\ap\ra\infty$ only acts on the latter and is given in \req{Limits}.
The result \req{result} generalizes in a geometric way  the EYM relations of \cite{ST,p} to any multiplicity  in terms of intersection numbers.

Let us present a simple example of \req{result} with $n=3,\;r=1$, i.e. $m=4$. In this case we have 
\be
PT^\vee(1,2,3,4)=-(p_1+p_2)^2\ \fc{d\mu_4}{\z_{12}\z_{24}\z_{43}\z_{31}}\equiv -(p_1+p_2)^2\ PT(1,2,4,3)\ ,
\ee
and from \req{TWISTEYMN1} we read off: 
\be
\lim_{\ap \rightarrow \infty} \widetilde{\varphi}^{EYM}_{+,n;r}=\fc{d\mu_4}{\z_{12}\z_{23}\z_{31}} 
\lf(\sum_{j=1}^3\fc{\eps_4p_j}{\z_4-\z_j}\ri)=(\eps_4p_1)\ PT(1,2,3,4)+(\eps_4p_2)\ PT(1,2,4,3)\ .
\ee
Eventually, with the intersection numbers  $\vev{PT(1,2,4,3),PT(1,2,3,4)}_\om=-\tfrac{1}{(p_1+p_2)^2}$
and $\vev{PT(1,2,4,3),PT(1,2,4,3)}_\om=\tfrac{1}{(p_1+p_2)^2}+\tfrac{1}{(p_1+p_3)^2}$ we find
\be
\Ac(3,1)=\lf\{(\eps_4p_1)+\fc{(p_1+p_4)^2}{(p_1+p_3)^2}\ (\eps_4p_2)\ri\}\ \Ac_{YM}(1,2,3,4)\ ,
\ee
in agreement with \cite{ST}.

Finally, let us compare the decomposition \req{result}   with the double--copy expression $EYM\!=\!YM\!+\!\phi^3\otimes YM$ \cite{Cachazo:2014xea} 
\be
\mathcal{A}(n;r)=\sum_{a,b=1}^{(m-3)!} \Ac_{gYMS}(b)\ S[b|a]\ \Ac_{YM}(a)\ ,
\ee
with the  KLT bilinear $S[b|a]$ and the $m$--point (single--trace) doubly--partial subamplitude $\Ac_{gYMS}(b)$ of generalized Yang--Mills--Scalar theory (gYMS) (pure YM plus cubic scalar theory) involving  $n$ scalars and $r$ gluons. This gives rise to  the relation
\be
\lim_{\ap \rightarrow \infty} \ \vev{PT^\vee(a), \widetilde{\varphi}^{EYM}_{+,n;r} }_\om=\sum_{b=1}^{(m-3)!}\Ac_{gYMS}(b)\ S[b|a]\ ,
\ee
which in turn expresses the intersection number in terms of gYMS theory.  A comprehensive extension of the
relations presented in this subsection will be performed elsewhere.

\goodbreak
\section{Concluding remarks}

In this work we have derived all--multiplicity expressions for EYM amplitudes in terms of 
twisted intersection numbers.
We have used the mixed superstring disk amplitude involving both open and closed strings to  find a pair of twisted differentials. Their intersection numbers  reproduce the CHY formula \req{AmpEYM} for EYM amplitudes in the infinite inverse tension limit $\ap\ra\infty$.
Our expressions for the twisted differentials are very compact and provide 
results \req{FinalAmp} for any multi--leg case.
Twisted differentials are naturally defined on the Riemann sphere. To find the latter from the disk we have established  a map \req{MAP} from the disk onto the sphere. This map converts the mixed superstring disk amplitude to a KLT like expression with factorized holomorphic and anti--holomorphic pieces with each piece giving rise to the relevant twisted differential form.
Our embedding from the disk to the sphere resembles the construction of heterotic string theory in which one has a superstring sector for right--movers and a bosonic string sector for the left--movers. This way our map links EYM amplitudes built from superstring disk amplitudes  \cite{Stieberger:2014cea,Stieberger:2015qja,Stieberger:2015kia,Stieberger:2015vya,ST} to EYM amplitudes engineered from closed string heterotic amplitudes  \cite{Bern:1999bx,Schlotterer:2016cxa}. 

It would be interesting to find connections between our construction and that of the  
heterotic ambitwistor string theory.
In fact, it is worth comparing our findings with the ambitwistor  results in \cite{Mason}. In  ambitwistor  string theory only the massless part of the spectrum survives and it is used to compute the CHY formula for gauge and gravitational amplitudes. In contrast, here we use in the string path integral that part of perturbative string theory,  which is dominant in the  high energy limit $\alpha' \!\rightarrow\! \infty$ to construct the appropriate pair of twisted differential forms \req{nrtwistedgen1} and \req{nrtwistedgen2}. This procedure
not only provides the localization  of the integrand over the solutions of the scattering equations but also eliminates the double poles associated to the  contraction of 
bosonic fields $\partial X$. On the other hand, in the ambitwistor case the path integral over the fields $X$ imposes the localization over solutions to the scattering equations. We  may interpret  our embedding \req{MAP} from the disk onto the sphere as a geometrical map from perturbative string theory to ambitwistor string theory.

We have presented the  all--multiplicity relation \req{result} expressing any
EYM amplitude involving $n$ gluons and $r$ gravitons as a linear combination of a $(m-3)!$ dimensional basis of pure $m$ gluon subamplitudes, with $m\!=\!n\!+\!r$. These findings, which simply follow from \req{FinalAmp}, generalize the results of \cite{ST,p} towards a geometric interpretation by twisted intersection theory. 

Generalizing the EYM tree--level relations \cite{Stieberger:2014cea,Stieberger:2015qja,Stieberger:2015kia,Stieberger:2015vya,ST} to one--loop is of considerable interest and can be accomplished by using the results
of \cite{Stieberger:2021daa}. A subsequent question is finding appropriate  twisted forms on the elliptic curve which generalize \req{Witten} and give rise to one--loop integrands of gauge and gravity theories. The one--loop monodromy results  from \cite{Hohenegger:2017kqy,Casali:2019ihm,Stieberger:2021daa} are of importance  for constructing such differential forms.

\appendix

\section{Unifying world--sheet description of gluons and gravitons}
\label{Notation}

In this appendix  we  introduce a universal notation for the  momenta, polarizations and positions of the string vertex operators on the world--sheet. This dictionary unifies the appearance of gluons and gravitons in our construction for the twisted differential forms.

There are two different types of momenta. One set of momenta $k_i$ refers to   the open string (gluon) legs while the set of momenta $q_i,\tilde q_i$ relates to closed string (graviton) legs. Since we are restricting to Neumann boundary conditions we can choose $q_i=\tilde q_i$. These two different sets are unified into a larger set of momenta denoted by $p_i$ and described by:
\begin{equation}
p:=\Big\{\{k_i\}_{i=1}^n \cup \{q_i,\widetilde{q}_i\}_{i=1}^r\Big\}\ .
\end{equation}
As an example for the case of three gluons and two gravitons we are dealing with the  following set of seven momenta $p$:
 \begin{equation}
p=\Big\{k_1,k_2,k_3,q_1,\widetilde{q}_1,q_2,\widetilde{q}_2\Big\} \label{toy}\ .
 \end{equation}
For generic $n$ and $r$ one can relate the elements of the set $p$ to the graviton and gluon momenta as:
\begin{equation}
p_m=\begin{cases}
&  m>n \,\,\, \begin{cases} \textit{if}\,\,\,{\scriptstyle{m=n+2l-1}}\,\, \,\,\, p_m=q_{l},  \\
  \textit{if}\,\,\,{\scriptstyle{m=n+2l}} \hspace{.65cm} p_m=\widetilde{q}_{l},   \end{cases} \\
 &  m \leq n  \hspace{1cm}p_{m}=k_m,
\end{cases}\label{P}
\end{equation}
E.g. for (\ref{toy})  we have:
 $$\begin{aligned}
& p_6=p_{3+2(2)-1}=q_2\ , \\
&  p_5=p_{3+2(1)}=\widetilde{q}_1\ .
\end{aligned}
$$
Similarly, for the gluon $\varepsilon_i$ and graviton polarizations  $\varepsilon_i, \widetilde{\varepsilon}_i$  we use the unifying set $\xi_i$ as follows:
\begin{equation}
\xi_i:=\Big\{\{\varepsilon_i\}_{i\leq n} \cup \{\varepsilon_i,\widetilde{\varepsilon}_i\}_{> n}\Big\}\ .
\end{equation}
For the previous example of three gluons and two gravitons we then deal with the following set $\xi$:
 \begin{equation}
\xi=\{\varepsilon_1,\varepsilon_2,\varepsilon_3,\varepsilon_4,\widetilde{\varepsilon}_4,\varepsilon_5,\widetilde{\varepsilon}_5\}\ .\label{toye}
 \end{equation}
Furthermore, we get the relation for the elements of the set $\xi$ to the  $n$ gluon and $r$ graviton polarizations:
\begin{equation}
\xi_m=\begin{cases}
&  m>n \,\,\, \begin{cases} \textit{if}\,\,\,{\scriptstyle{m=n+2l-1}}\,\, \,\,\, \xi_m=\varepsilon_{l+n}, \\
  \textit{if}\,\,\,{\scriptstyle{m=n+2l}} \hspace{.65cm} \xi_m=\widetilde{\varepsilon}_{l+n},   \end{cases} \\
 &  m\leq n  \hspace{1cm}\xi_{m}=\varepsilon_m\ .
\end{cases}\label{pol}
\end{equation}
Again, for the previous example  (\ref{toye})  we have: $$\begin{aligned}
& \xi_6=\xi_{3+2(2)-1}=\varepsilon_5\ , \\
&  \xi_5=\xi_{3+2(1)}=\widetilde{\varepsilon}_4\ .
\end{aligned}
$$
 Finally, to use a unifying description for the string vertex operator positions 
 we introduce  $\zeta_i$ accounting for the open string positions $(\{z_i\}_{i\leq n})$ and the closed string positions $(\{z_i\}_{i>n})$, respectively:
\begin{equation}
\zeta_i:=\Big\{\{z_i\}_{i\leq n} \cup \{z_i,z_i\}_{i> n}\Big\}\ .
\end{equation}
Where $n$ is the number of gluons and $i$ index runs up to $n+r$.\\
Following our three gluon and two graviton example, it is characterized  by the following set:
 \begin{equation}
\zeta=\{z_1,z_2,z_3,z_4,z_4,z_5,z_5\}\ . \label{toyz}
 \end{equation}
The elements of the generic set $\zeta$ describing $n$ gluons and $r$ gravitons are:
\begin{equation}
\zeta_m=\begin{cases}
&  m>n \,\,\, \begin{cases} \textit{if}\,\,\,{\scriptstyle{m=n+2l-1}}\,\, \,\,\, \zeta_m=z_{l+n}, \\
  \textit{if}\,\,\,{\scriptstyle{m=n+2l}} \hspace{.65cm} \zeta_m=z_{l+n},   \end{cases} \\
 &  m\leq n  \hspace{1cm}\zeta_{m}=z_m\ ,
\end{cases}\label{zeta}
\end{equation}
which in turn gives for our example (\ref{toyz}):
 $$\begin{aligned}
& \zeta_6=\zeta_{3+2(2)-1}=z_5\ , \\
&  \zeta_5=\zeta_{3+2(1)}=z_4\ .
\end{aligned}
$$
As we can see the set $\zeta$ is a multiple set i.e. there exist duplicate elements for closed strings. We are going to see in the next sections that this choice helps us write the general amplitude in a concise way.

\section[Calculation of correlator ${\bf \psi_{1}}$] {Calculation of correlator $\psi_{1}$}\label{A1}

Here we discuss how the factor $C_{qq}$ appearing in CHY formalism \req{cqq} is reproduced in the superstring amplitude. For this we need to consider  the contractions of $\partial X^{\mu}(z) $ (with $z$ denoting the dependence on the graviton vertex position) with the exponential factor  of the gluons $\prod\limits_{i=1}^n e^{ik_i X^\nu_i (x_i)}$:
\begin{equation}
    \begin{aligned}
     & \psi_{1}=\langle \varepsilon_\mu \partial X^{\mu}(z)  \prod\limits_{i=1}^n e^{ik_i X^{\nu_i} (x_i)} \rangle=i\sum\limits_{i=1}^{n} \frac{\varepsilon \cdot k_i}{x_i-z}=i\sum\limits_{i=1}^{n-1} \frac{\varepsilon \cdot k_i}{x_i-z}+i \frac{\varepsilon \cdot k_n}{x_n-z}\\
     & =i\sum\limits_{i=1}^{n-1} \frac{\varepsilon \cdot k_i}{x_i-z}-i \frac{\varepsilon \cdot \sum_{j=1}^{n-1}k_j}{x_n-z}=i \sum\limits_{i=1}^{n-1}(\varepsilon \cdot k_i) \frac{(x_{ni})}{(x_i-z)(x_n-z)}\\
     & =i \sum\limits_{i=1}^{n-1}(\varepsilon \cdot k_i) \sum_{l=i}^{n-1} \Bigg(\frac{1}{x_l-z}-\frac{1}{x_{l+1}-z}\Bigg)=-i \sum\limits_{i=1}^{n-1}(\varepsilon \cdot k_i) \sum_{l=i}^{n-1} \frac{x_{l,l+1}}{(x_l-z)(x_{l+1}-z)}\\
     & =-i \sum\limits_{l=1}^{n-1}\Bigg[(\varepsilon \cdot \sum\limits_{j=1}^l k_j)  \frac{x_{l,l+1}}{(x_l-z)(x_{l+1}-z)}\Bigg]=-i \sum\limits_{l=1}^{n-1}(\varepsilon \cdot x_l)  \frac{x_{l,l+1}}{(x_l-z)(x_{l+1}-z)}\ . \label{Psi}
    \end{aligned}
\end{equation}
Here we have used the telescopic series summation and the definition $x_i=\sum_{j=1}^i k_j$. This demonstrates that the expression $C_{pp}$ from \req{cqq} originates from specific contractions in the superstring amplitudes.

\newpage
\addcontentsline{toc}{section}{References}

\bibliography{bibt}
\end{document}